\documentclass[nonacm,sigplan,10pt]{acmart}
\renewcommand\footnotetextcopyrightpermission[1]{}
\setcopyright{none}

\usepackage[]{hyperref}
\usepackage[english]{babel}
\usepackage{blindtext}
\usepackage{booktabs}
\usepackage{subcaption}
\usepackage{makecell}
\usepackage{soul}
\usepackage{times}
\usepackage{graphicx}
\usepackage{graphbox}
\usepackage{bm}
\usepackage{physics}
\usepackage{tikz}
\usetikzlibrary{arrows.meta,positioning,calc, patterns,fit,decorations.pathreplacing}

\usepackage{booktabs, multirow, colortbl, pifont, rotating}
\usepackage{makecell}

\usepackage{multicol}
\usepackage{caption}

\hypersetup{breaklinks=true}

\AtBeginDocument{%
  \urlstyle{same}%
}


\usepackage{listings}
\usepackage{xcolor}

\lstdefinelanguage{json}{
  showstringspaces=false,
  string=[s]"",
  morecomment=[l]{//},
}
\usepackage{multirow}
\usepackage{pgfplotstable}
\usepackage{tabularx, booktabs}

\usepackage{tikz}
\usetikzlibrary{calc,arrows.meta,positioning,decorations.pathreplacing}

\usepackage{pifont} 
\usepackage{longtable}
\usepackage{amsmath}
\usepackage{amsthm}

\usepackage{}

\usepackage{enumitem}

\usepackage{scalerel,stackengine}
\stackMath
\newcommand\rwcheck[1]{%
\savestack{\tmpbox}{\stretchto{%
  \scaleto{%
    \scalerel*[\widthof{\ensuremath{#1}}]{\kern-.6pt\bigwedge\kern-.6pt}%
    {\rule[-\textheight/2]{1ex}{\textheight}}
  }{\textheight}%
}{0.5ex}}%
\stackon[1pt]{#1}{\scalebox{-1}{\tmpbox}}%
}
\newcommand\rwhat[1]{%
\savestack{\tmpbox}{\stretchto{%
  \scaleto{%
    \scalerel*[\widthof{\ensuremath{#1}}]{\kern-.6pt\bigwedge\kern-.6pt}%
    {\rule[-\textheight/2]{1ex}{\textheight}}
  }{\textheight}%
}{0.5ex}}%
\stackon[1pt]{#1}{\tmpbox}%
}
\definecolor{darkred}{rgb}{0.55,0.0,0.0}
\definecolor{darkgreen}{rgb}{0.0,0.35,0.0}

\usepackage{xcolor}
\newcommand*\colourcheck[1]{%
  \expandafter\newcommand\csname #1check\endcsname{\textcolor{#1}{\ding{52}}}%
}
\colourcheck{darkgreen}
\newcommand*\colourcross[1]{%
  \expandafter\newcommand\csname #1cross\endcsname{\textcolor{#1}{\ding{56}}}%
}
\colourcross{darkred}

\usepackage{enumitem}

\newcommand{\xref}[1]{\S\ref{#1}}

\usepackage{array}
\newcolumntype{C}[1]{>{\centering\arraybackslash}p{#1}}

\newcommand{\squishlist}{
  \begin{list}{$\bullet$}{
    \setlength{\itemsep}{0pt}       \setlength{\parsep}{3pt}
    \setlength{\topsep}{3pt}        \setlength{\partopsep}{0pt}
    \setlength{\itemindent}{1em}
    \setlength{\leftmargin}{0em}    \setlength{\labelwidth}{1em}
    \setlength{\labelsep}{0.5em} } }

\newcommand{\squishend}{
\end{list} }

\newcounter{mycounter}

\usepackage{algorithm}
\usepackage{algpseudocode}

\newcommand{\projectname}{{\textsc{CacheWise}}}
\newcommand{\system}{\projectname}

\newcommand\tpu{{TPU}}
\newcommand\gpu{{GPU}}

\newcommand\xpu{{XPU}}
\newcommand\kvcache{{KVCache}}

\newcommand\codingagent{coding agent}

\newcommand\catraces{CATraces}

\begin{document}



\title{CacheWise: Understanding Workloads and Optimizing {\kvcache{}} Management for Efficiently Serving LLM Coding Agents}

\author{Shubham Tiwari}
\affiliation{%
  \institution{University of Washington}
  \city{Seattle}
  \state{WA}
  \country{USA}
}

\author{Tapan Chugh}
\affiliation{%
  \institution{University of Washington}
  \city{Seattle}
  \state{WA}
  \country{USA}
}

\author{Nash Rickert}
\affiliation{%
  \institution{University of Washington}
  \city{Seattle}
  \state{WA}
  \country{USA}
}

\author{Simon Peter}
\affiliation{%
  \institution{University of Washington}
  \city{Seattle}
  \state{WA}
  \country{USA}
}

\author{Ratul Mahajan}
\affiliation{%
  \institution{University of Washington}
  \city{Seattle}
  \state{WA}
  \country{USA}
}

\author{Haiying Shen}
\affiliation{%
  \institution{University of Virginia}
  \city{Charlottesville}
  \state{VA}
  \country{USA}
}

\renewcommand{\shortauthors}{}


\begin{abstract}
Coding agents are a fast-growing LLM application, executing as long-running closed-loop sessions in which LLM generations alternate with external tool calls. Yet, unlike chat workloads, their serving behavior has not been studied extensively. We address this gap by collecting a dataset of real-world coding assistant traces. Our analysis shows that coding agent sessions repeatedly reuse large prefixes and create sustained {\kvcache{}} pressure that conventional LLM serving policies handle poorly.

Based on our analysis, we present {\projectname}, a {\kvcache{}} management layer that improves {\kvcache{}} reuse for coding agent workloads. {\projectname} combines prefix-aware scheduling with reuse-aware eviction guided by lightweight predictions from tool call metadata. Implemented in vLLM and evaluated on the collected traces, {\projectname} reduces {\kvcache{}} evictions by up to $2$--$2.6\times$ and improves total agent session completion time by up to \textasciitilde$3.5\times$. \projectname{} is open sourced at \href{https://github.com/cachewise-project/cachewise-coding-traces}{github.com/cachewise-project/cachewise-coding-traces}
\end{abstract}
\settopmatter{printfolios=true, printacmref=false}
\maketitle

\pagestyle{plain}

\section{Introduction}

\begin{figure}[t]
\centering
\begin{subfigure}[t]{\columnwidth}
    \centering
\includegraphics[width=0.7\columnwidth]{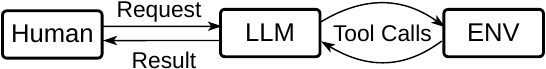}
    \caption{Coding agent workflow.}
    \label{f:closed_loop_session}
\end{subfigure}

\begin{subfigure}[t]{0.49\columnwidth}
    \centering
    \includegraphics[width=\linewidth]{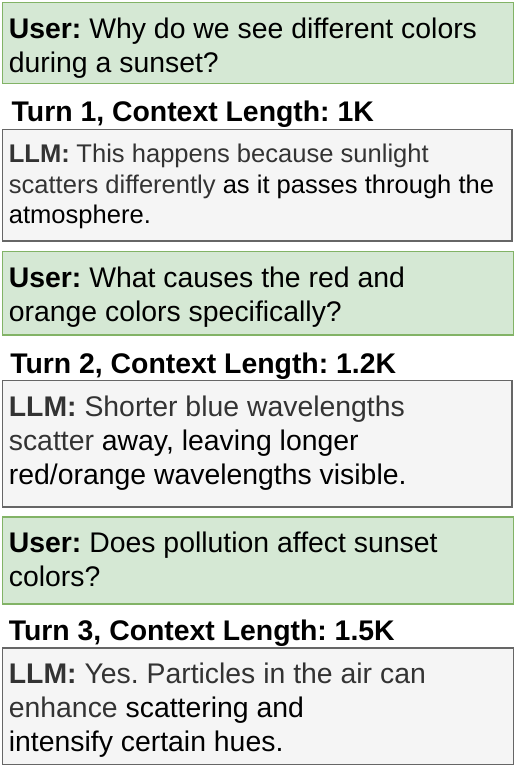}
    \caption{Chat session.}\label{fig:chat-trace}
\end{subfigure}
\hfill
\begin{subfigure}[t]{0.49\columnwidth}
    \centering
    \includegraphics[width=\linewidth]{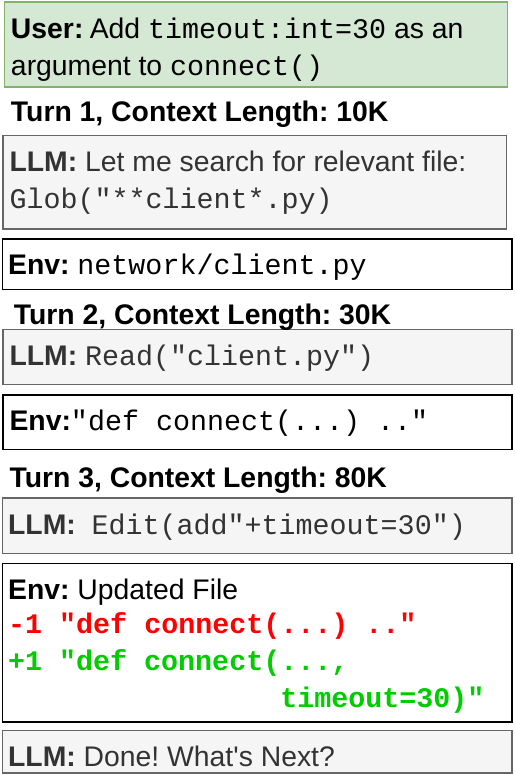}
    \caption{Coding assistant session.}
    \label{fig:coding-trace}
\end{subfigure}
\caption{Chat and coding-assistant interactions.}
\label{fig:example-traces}
\end{figure}

Coding assistance~\cite{claude_code, openai_codex, github_copilot}, where a Large Language Model (LLM) generates code based on user instructions, is emerging as a crucial application with significant potential. Unlike chatbots (\autoref{fig:chat-trace}), where the user sends a query and the LLM generates a corresponding response, coding agents~\cite{yang2024sweagent, zhang2024autocoderover, jimenez2024swebench} formulate multi-step plans and execute them through a series of tool calls~\cite{gorilla, yao2023react} to run code, inspect program outputs, or modify files (\autoref{fig:coding-trace}). As \autoref{f:closed_loop_session} shows, the execution alternates between the LLM, which processes input and generates function calls, and the \emph{environment}, which executes those calls and performs actions such as reading/writing code and running tests.

While LLM serving in the context of chat systems has been studied extensively~\cite{vllm, sglang, mooncake, distserve, orca, splitwise, sarathi-serve, sun2024llumnix, fu2024serverlessllm}, \codingagent{} serving systems have not been examined previously, to the best of our knowledge. A key reason is the lack of publicly available real-world traces for this workload. To address this gap, we collect a dataset of real-world coding assistant traces from researchers actively using coding assistants for research. 

In this work, we use this collected dataset to study coding agent workloads and their implications for serving system design. Compared to chatbot workloads, coding-agent sessions are long-running, accumulate substantially larger contexts, and are dominated by tool-initiated turns rather than direct user input. A single user task therefore expands into a closed-loop sequence of LLM requests and tool calls, where each new turn reuses and extends a growing prefix. Consequently, coding assistance is fundamentally a \emph{session-oriented} serving workload: the key objective is not the latency of an individual request, but the completion time of the full multi-turn session.

Existing LLM serving systems are primarily designed and evaluated for prior workloads such as chatbot applications. Their default scheduling and {\kvcache} management policies are not well-suited for \codingagent{} workloads. In particular, systems such as vLLM~\cite{vllm} and Mooncake~\cite{mooncake} employ First-Come-First-Served (FCFS) batching and Least-Recently-Used (LRU) {\kvcache} eviction. These workload-agnostic policies ignore two defining properties of coding-agent sessions: requests from the same session frequently have high prefix overlap with already-resident state, and the time to next reuse depends strongly on the ongoing tool execution. FCFS interleaves many sessions and expands the active {\kvcache} working set, increasing the likelihood of evicting prefixes that will soon be needed again. LRU, in turn, uses only past access recency and cannot distinguish between a session whose tool call is about to complete and one whose state will remain unused for much longer. Under memory pressure, these policies therefore trigger {\kvcache} thrashing, reducing useful work and lowering token goodput (i.e., the rate of useful token processing after excluding recomputation and {\kvcache} movement overheads). While some recent works~\cite{gim2025pie,autellix} attempt to optimize serving for agentic workloads, they tightly couple the agent implementation framework with the LLM serving system, and thus are ill-suited for large-scale deployments, which must support diverse clients.

To address these challenges, we introduce {\projectname}, an agent-aware {\kvcache} management layer that reduces the overhead from {\kvcache} recomputation and offload for general-purpose LLM serving systems. {\projectname} is built on two key ideas: prefix-aware request scheduling and reuse-aware {\kvcache} eviction. Prefix-aware request scheduling prioritizes inference requests with a higher degree of prefix overlap with {\kvcache} state already maintained in accelerator memory. By doing so, it reduces the number of blocks that must be reclaimed and later restored. Complementing this, reuse-aware {\kvcache} eviction evicts prefixes based on anticipated reuse likelihood, rather than relying purely on access recency. Notably, {\projectname} leverages session metadata, such as tool execution information, as signals to predict the reuse likelihood of {\kvcache} prefixes using lightweight predictors trained on historical samples collected by the serving system, without any modifications to the coding agents themselves. This enables, {\projectname} to sustain high accelerator utilization while reducing the frequency of {\kvcache} evictions, thereby improving overall token goodput, i.e., the rate at which new LLM tokens are generated.

This paper makes the following contributions:
\squishlist
    \item We collect and analyze real-world \codingagent{} traces, and use them to characterize \codingagent{}s as a distinct class of LLM serving workloads. We contrast them with traditional chatbot and other multi-turn tool-use workloads, identify the systems implications of closed-loop execution, long-lived sessions, and large, growing prefixes. We have open sourced the dataset at \href{https://github.com/cachewise-project/cachewise-coding-traces}{github.com/cachewise-project/cachewise-coding-traces}.

    \item We design and implement {\projectname}, based on two key ideas: prefix-aware scheduling and predictive {\kvcache} eviction. We implement our techniques on top of vLLM~\cite{vllm}.

    \item We evaluate our techniques using real-world \codingagent{} traces. Our results show that {\projectname} reduces {\kvcache} evictions by $2$--$2.6\times$ compared to state-of-the-art systems such as vLLM, significantly improving token goodput. In addition, {\projectname} improves total request completion time by up to \textasciitilde$3.5\times$ and achieves performance comparable to a predictive eviction policy with ground-truth reuse information.
\squishend


\section{Background: Efficient LLM Serving}
\label{sec:background}
\label{sec:background-llm-serving}

LLM serving typically requires expensive and scarce accelerators such as \gpu{}s and \tpu{}s (we will collectively call them {\xpu}s). Efficiently using their massive internal resources requires massively parallel operations to optimize their internal compute and memory resources. Although serving requests consist of separate \textit{prefill} and \textit{decode} stages, which require fundamentally different optimization strategies, efficient {\xpu} memory management is crucial for both.

During prefill, serving requires a large number of parallel compute operations to process the input sequence and generate the intermediate key-value tensors (KVCache) corresponding to each input token. During decode, efficient serving systems typically employ \textit{continuous batching}~\cite{orca,vllm} to ensure parallelism \emph{across requests}, since the LLM only generates output tokens sequentially, due to the autoregressive nature of the Transformer architecture~\cite{transformers}. While
parallelism within a sequence (for prefill) or across multiple sequences, through batching~\cite{orca} (for decode) improves {\xpu} efficiency, the benefits are limited by the {\xpu} memory available to store the {\kvcache} states of all the requests in the batch.

Similarly, the available {\kvcache} memory is a bottleneck for optimizing the prefill phase of serving requests that share a common prefix. Although state-of-the art serving systems, e.g., vLLM~\cite{vllm}, SGLang~\cite{sglang}, persist {\kvcache} from previous requests to minimize the number of operations required for executing requests that extend a previous request, during memory pressure, persisted {\kvcache} prefixes must be evicted to allocate memory for requests whose prefixes are not already available. If a subsequent request arrives whose prefix corresponds to evicted {\kvcache} state, the prefix must be rematerialized which either requires recomputing the {\kvcache} or transferring it back from lower storage tiers, such as host DRAM, SSD, or remote memory, if available (e.g., LMCache~\cite{lmcache}, Mooncake~\cite{mooncake}). Notice that although some systems, e.g., Infercept~\cite{infercept}, make {\kvcache} management decisions per-request, state-of-the-art serving systems typically implement a block-based abstraction, analogous to memory paging in operating systems, that partitions the {\kvcache} memory pool in the {\xpu} into fixed-size blocks, and eviction decisions can be made at per-block to reduce wastage.

\begin{table}[t]
\centering
\footnotesize{
  \begin{tabular}{r||C{0.30in}C{0.52in}C{0.15in}C{0.27in}C{0.35in}}
  \textbf{Trace} & \textbf{\catraces{}} & \textbf{SWE-} & \textbf{T1} & \textbf{Toucan} & \textbf{ShareGPT} \\
    \textbf{Captures?} & \textbf{(ours)} & \textbf{Agent}\cite{swe-smith} & \cite{T1} & \cite{toucan} & \cite{sharegpt52k} \\
  \hline
  \hline
  \# Tokens & 10M & 474M & 2.3K & 0.94M & 0.26M \\
  \hline
  Workload & Coding & Software & Tool & Tool & \multirow{2}*{Chat} \\
  Domain & Assistant & Engg. & Usage & Usage & \\
  \hline
  Interactive & \multirow{2}*{\darkgreencheck{}} & \multirow{2}*{\darkredcross{}} & \multirow{2}*{\darkredcross{}} & \multirow{2}*{\darkredcross{}} & \multirow{2}*{\darkgreencheck{}} \\
  Sessions? & & & & & \\
  \hline
  Real Tasks? & \darkgreencheck{} & \darkredcross{} & \darkredcross{} & \darkredcross{} & \darkgreencheck{} \\
  \hline
  Real Tools? & \darkgreencheck & \darkgreencheck & \darkredcross{} & \darkgreencheck{} & \darkredcross{} \\
\end{tabular}
}
\caption{Comparison of coding agent and multi-turn datasets.}
\label{tab:dataset_comparison}
\end{table}

\section{Coding Agent Workloads}\label{sec:workload-analysis}

To understand the system implications of \codingagent{} workloads, we present a detailed characterization of traces collected from users of Claude Code~\cite{claude_code}, a popular coding assistant application, performing real-world software development and engineering tasks on various open source projects.

Our collected dataset\footnote{We collected anonymized traces from consenting participants within our lab. This data collection follows standard ethical guidelines for research.}, denoted as \catraces{}, includes detailed conversations: user instructions, model outputs, tool calls, their results, etc., and annotated with key system metadata e.g,. number of tokens, timestamps for each message, human interventions, and so on.
Table~\ref{tab:dataset_comparison} highlights the unique aspects compared to previously available LLM workload traces. To the best of our knowledge, we are the first to present the real-world usage of coding assistant workloads; ShareGPT~\cite{sharegpt52k} presents real-world interactions for chat workloads, while others usually contain synthetic tasks~\cite{swe-smith,T1,jimenez2024swebench}, sometimes against mock tools~\cite{toucan}.

\begin{figure*}[t]
  \centering
  \begin{subfigure}{0.49\linewidth}
  \centering
  \includegraphics[width=\linewidth]{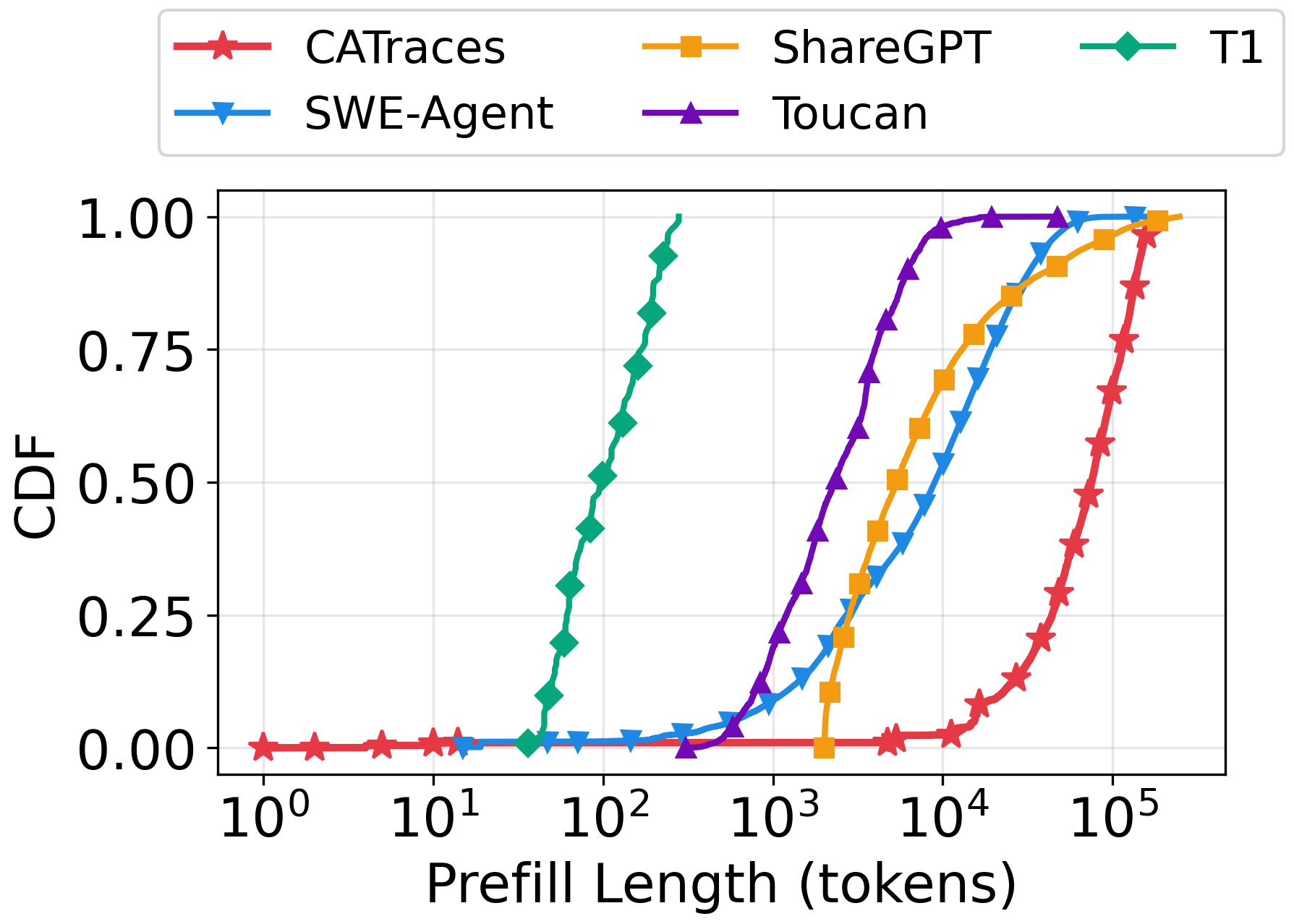}\caption{Prefill length distributions for different datasets}
  \end{subfigure}\hfill
  \begin{subfigure}{0.49\linewidth}
  \centering
  \includegraphics[width=\linewidth]{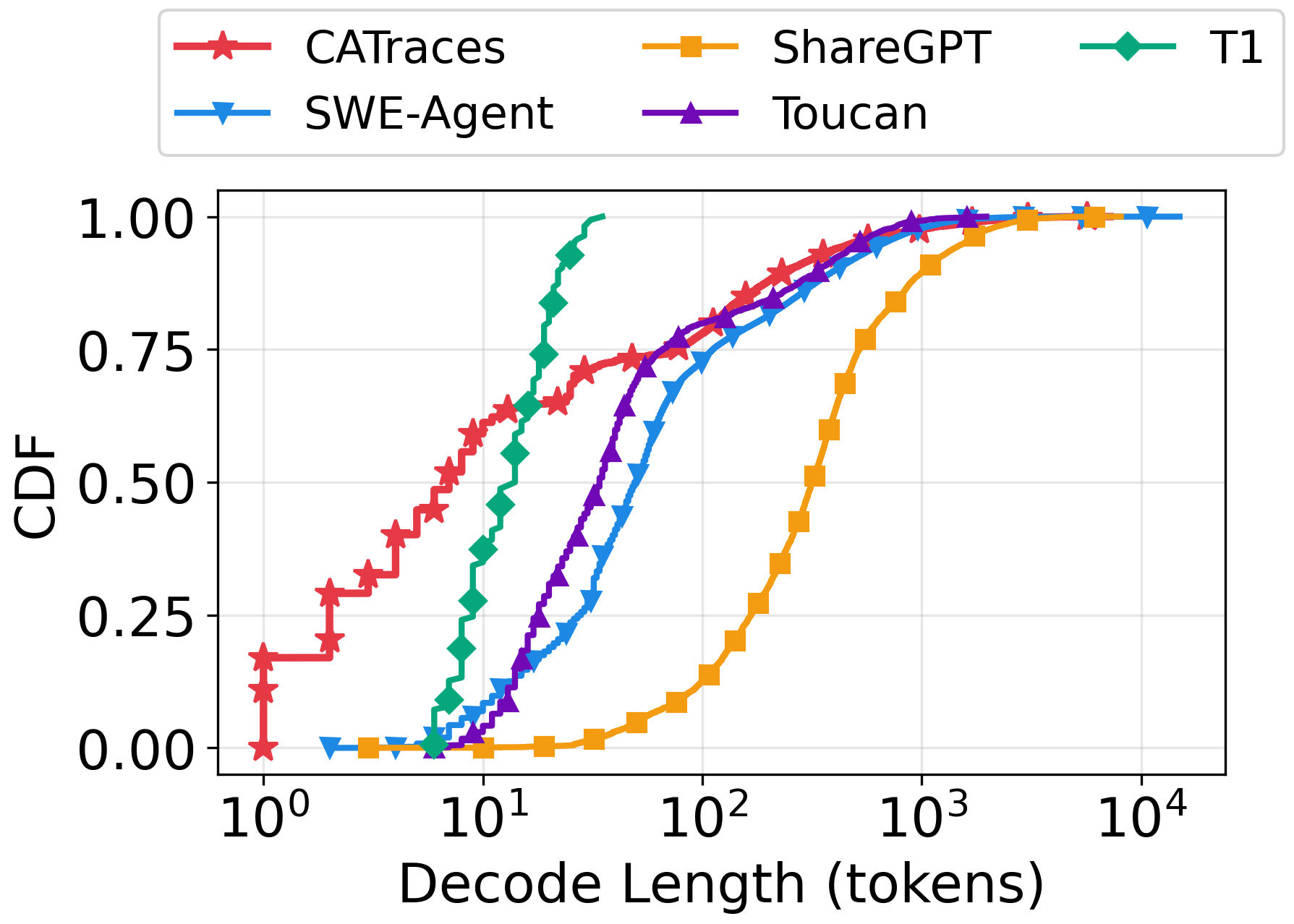}\caption{Decode length distributions for different datasets}
  \end{subfigure}  
  \caption{Comparing LLM serving request characteristics across datasets.}\label{fig:prefill-decode-cdfs}
\end{figure*}

\begin{figure}
  \centering
  \includegraphics[width=0.8\linewidth]{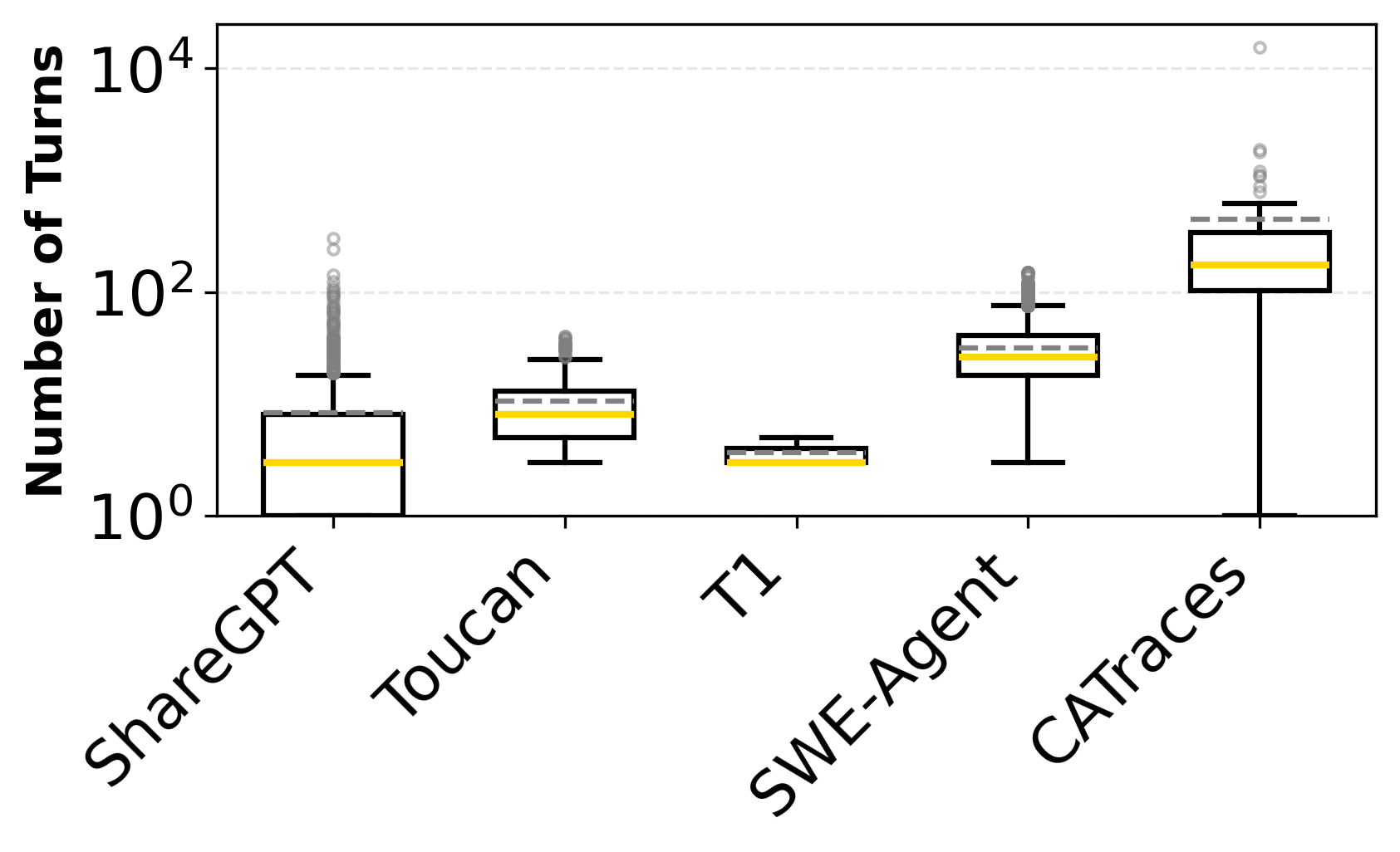}
  \caption{Distribution of number of turns for different workloads.}
  \label{fig:num_turns}
\end{figure}

\vspace{0.05in}
\noindent\textbf{\catraces{} Overview:} We begin by characterizing structural differences between real-world coding agent workloads and prior datasets. First, observe in Figure \ref{fig:num_turns}, that {\codingagent} serving requests have orders of magnitude more \emph{turns} than other workloads, where a turn refers to a new LLM request building upon a previous request and result. Notice that each turn presents an opportunity for {\kvcache} reuse; if the entire accumulated prefix needed to be recomputed for each turn, the overall number of flops required to compute would be orders to magnitude higher. Second, observe in \autoref{fig:prefill-decode-cdfs} that the distribution of the number of tokens to prefill and decode per request is significantly different for {\codingagent} workloads compared to others: without {\kvcache} sharing across turns, each request must prefill a substantially higher number of tokens, whereas each generates significantly fewer decode tokens, leading to \textasciitilde$21\times$ higher ratio of prefill to decode tokens compared to chatbot workloads. These differences have crucial implications for future systems~\cite{du2025prefillonly, zhang2025jenga}: the significantly larger shared {\kvcache} prefixes implies that only fewer \kvcache{} blocks can be stored in {\xpu} memory concurrently, which reduces the number of requests that can be executing concurrently, and constrains the overall token generation throughput during the decode phase (\xref{sec:background-llm-serving}). Overall, optimizing {\kvcache} management is especially crucial for serving {\codingagent} workloads efficiently.

\begin{figure}[h]
    \centering
    \includegraphics[width=0.9\linewidth]{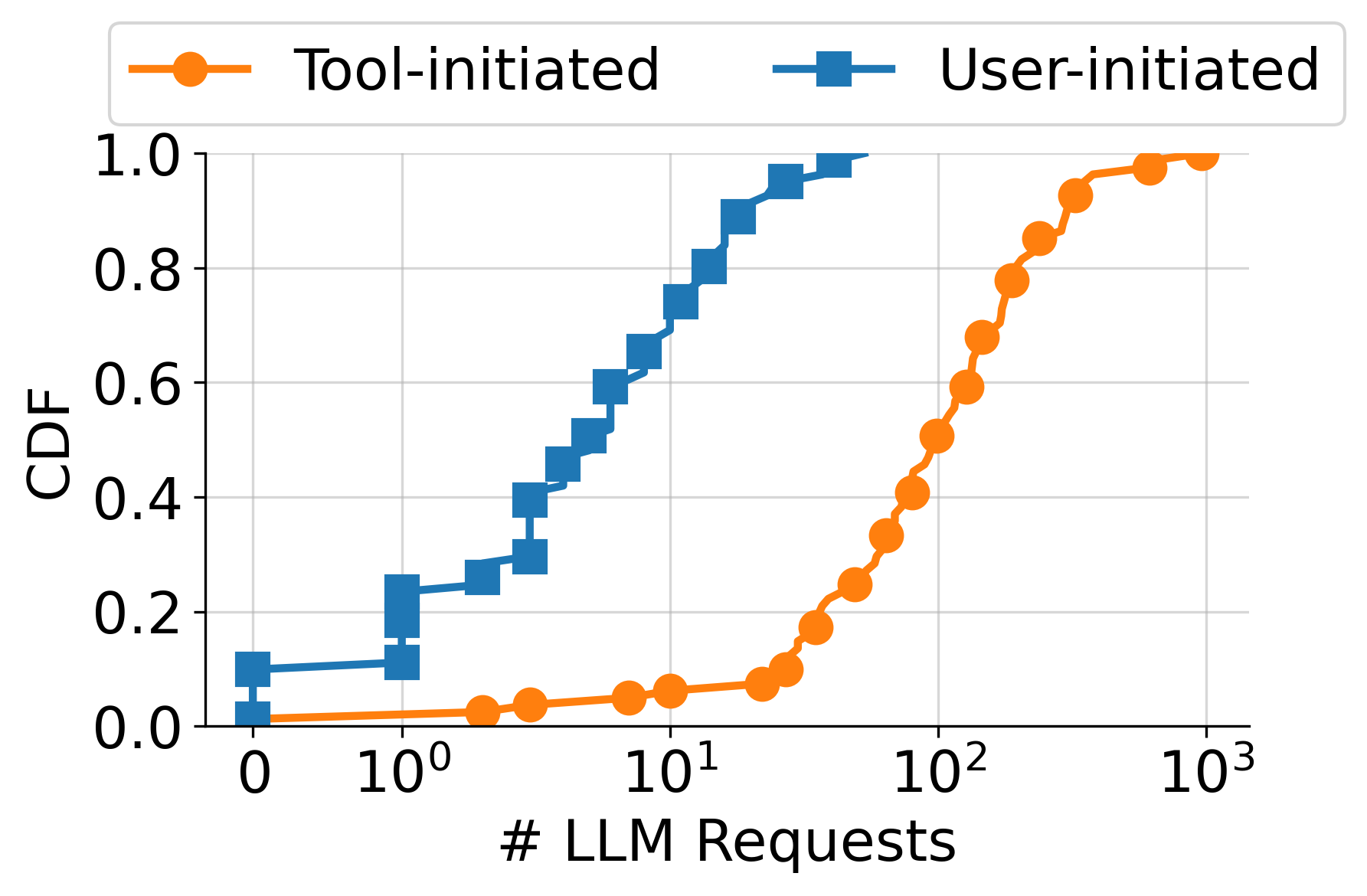}
    \caption{CDF of requests triggered by tool completion versus user input. Tool-initiated requests dominate, indicating a predominantly closed-loop request generation process.}
    \label{fig:tools_vs_users}
\end{figure}

\vspace{0.05in}
\noindent\textbf{Closed-loop execution shifts performance objective:}
We analyze the distribution of requests initiated by tool completion versus those initiated by direct user input. \autoref{fig:tools_vs_users} shows this distribution. We observe that requests triggered by tool completion (i.e., model-generated tool calls)~\cite{yao2023react, shinn2023reflexion} are $20\times$ more frequent than user-initiated requests at the median, indicating that the request generation process is predominantly closed-loop. 
\autoref{fig:example-traces} illustrates this behavior: while chat workloads consist of simple user–LLM interactions, \codingagent{} workloads form closed-loop chains where a single user request triggers a sequence of LLM requests and tool calls. To evaluate the performance of \codingagent{} workloads, session-level metrics (such as session completion time) capture the system’s efficiency in advancing multi-step execution within a session, whereas widely used token-level metrics such as Time-To-First-Token (TTFT) and Time-Between-Tokens (TBT) capture only per-request latency.

\begin{figure}[h]
    \centering
    \includegraphics[width=0.8\linewidth]{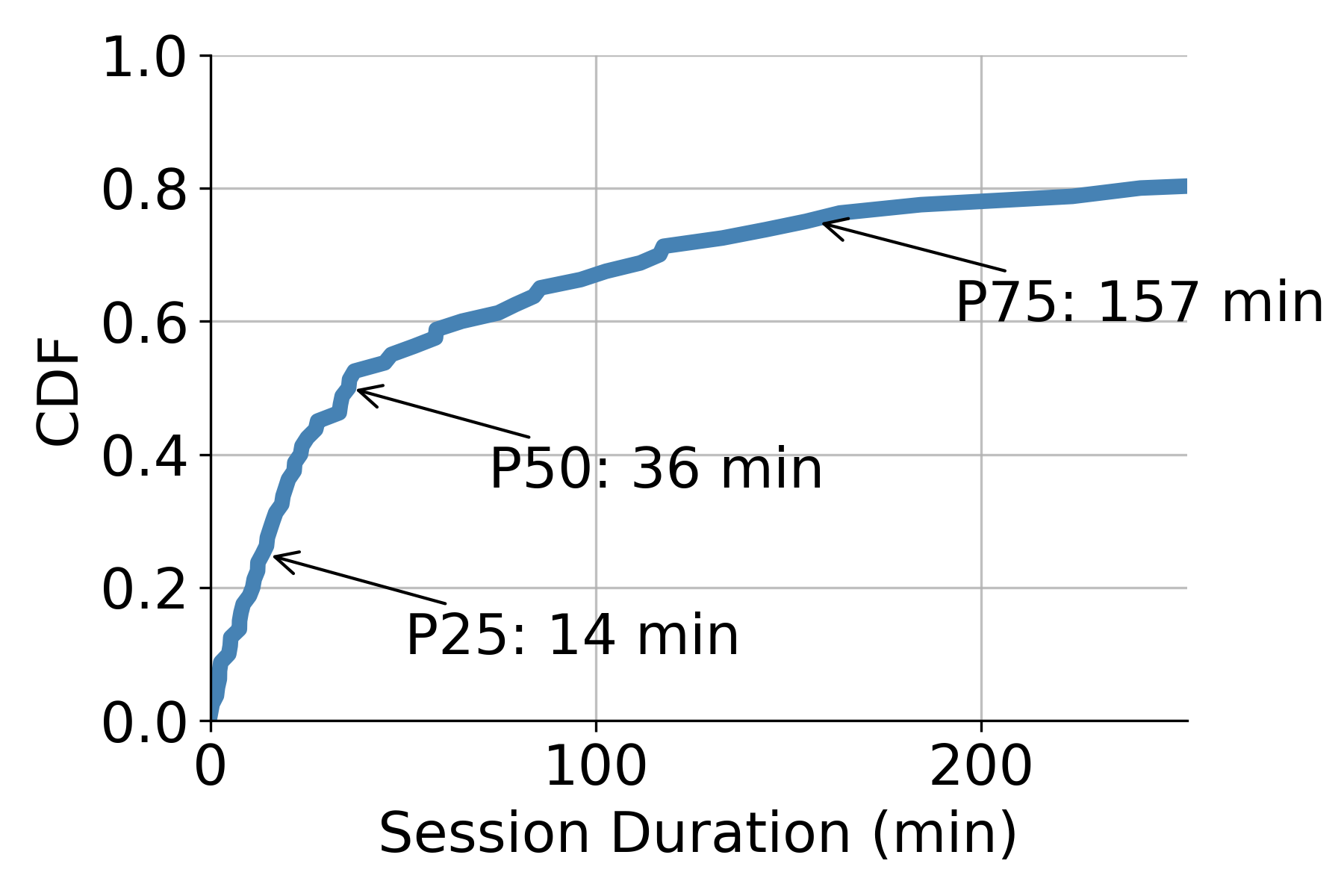}
    \caption{Session duration distribution observed in \catraces{}.}
    \label{fig:session_duration}
\end{figure}

\begin{figure}[h]
    \centering
    \includegraphics[width=0.8\linewidth]{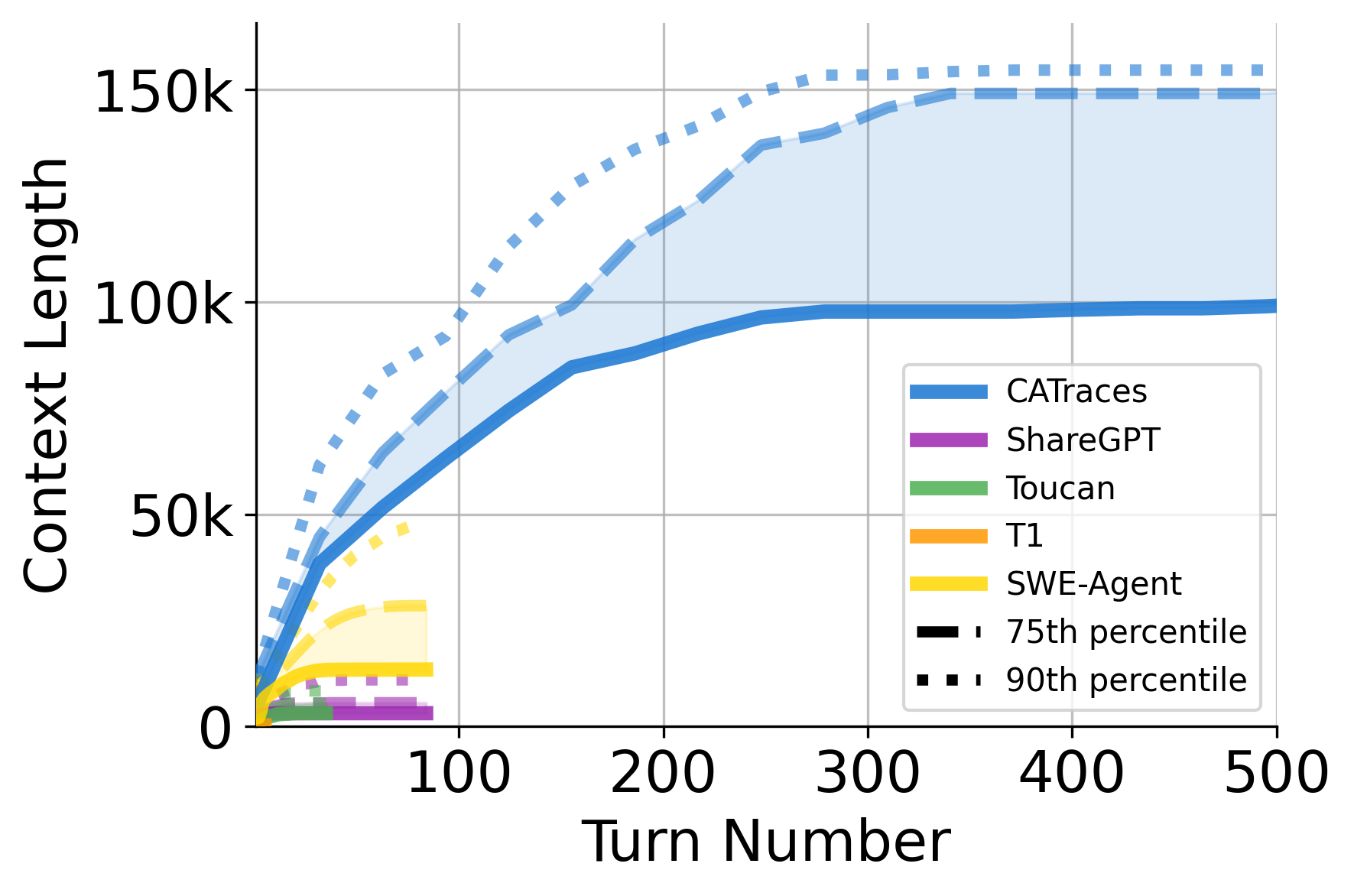}
    \caption{Context Length growth in number of tokens.}
    \label{fig:context_evolution}
\end{figure}

\vspace{0.05in}
\noindent\textbf{Long-running sessions compete for limited accelerator memory:}
We analyze session durations by computing the time difference between the first and last LLM request in each session of \catraces{}. 
\autoref{fig:session_duration} 
shows the distribution of session durations. We observe substantial variation in duration, with long-running sessions being common (e.g., $36$ min at median, >$2.6$ hours at tail). Such long-running sessions imply that \kvcache{} state must persist in \xpu{} memory over long periods to enable reuse across turns. Each turn of the session appends additional context to the prefix, causing the \kvcache{} memory to grow over time (as shown in \autoref{fig:context_evolution}), and remain resident for the duration of the session. As multiple long-running sessions coexist, their \kvcache{} state collectively competes for limited \xpu{} memory.
Large \kvcache{} prefixes directly translate to higher eviction overheads. Evicting large \kvcache{} states requires recomputation of state or data transfer across memory tiers, both of which incur significant overhead and impact system's efficiency. 

\begin{figure}[t]
    \centering
    \includegraphics[width=0.8\linewidth]{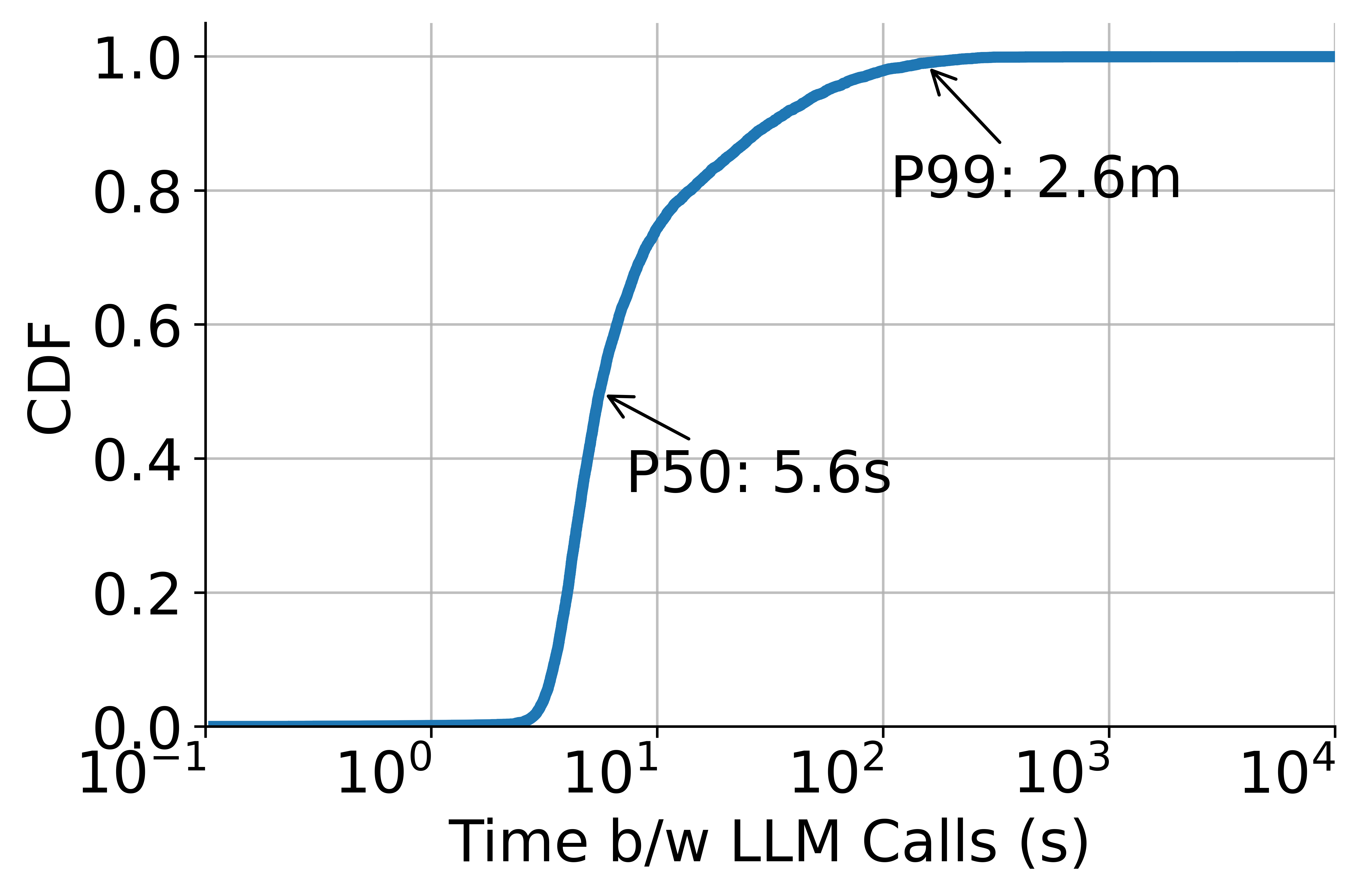}
    \caption{Time between consecutive LLM requests.}
    \label{fig:inter_llm_time}
\end{figure}

\begin{figure}[t]
    \centering
    \includegraphics[width=0.9\linewidth]{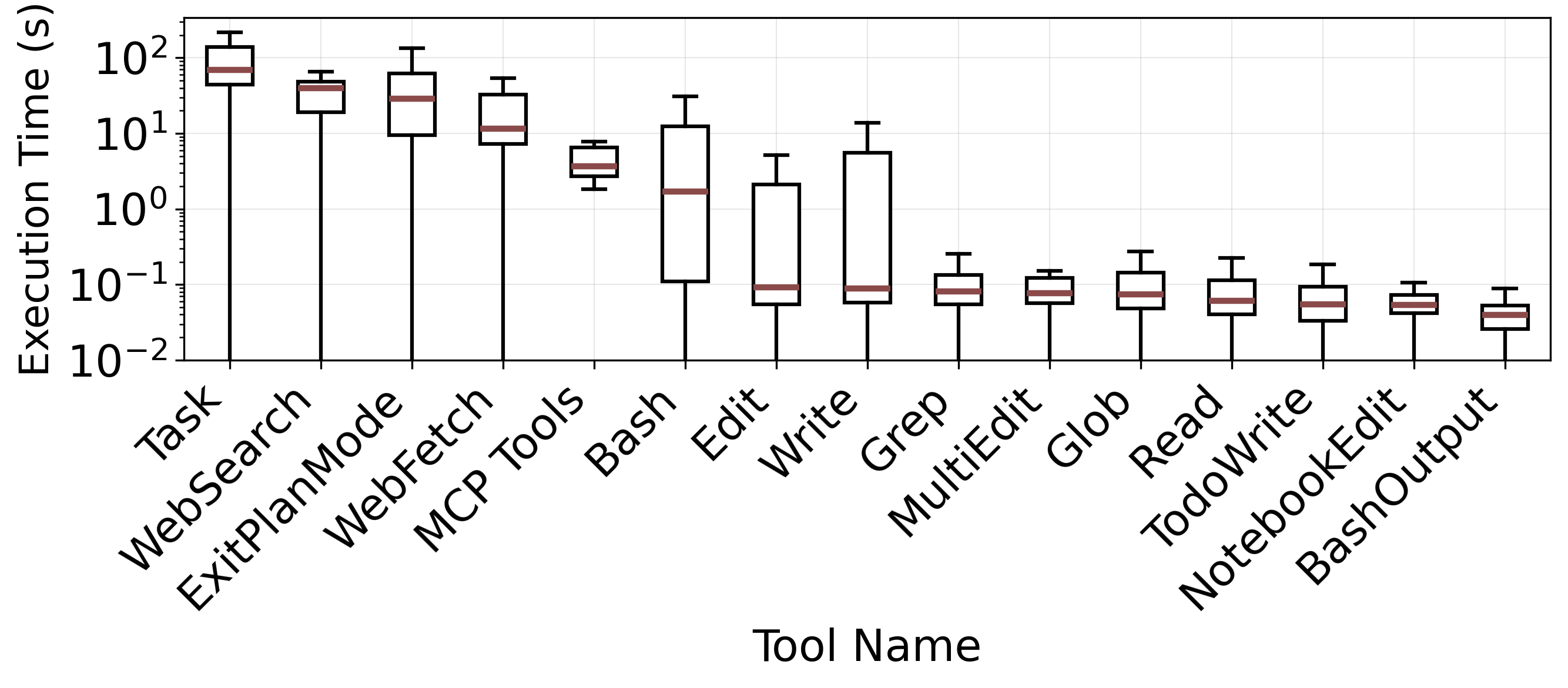}
    \caption{Distribution of tool execution durations.}
    \label{fig:tool_call_dist}
\end{figure}

\vspace{0.05in}
\noindent\textbf{Inter-request time varies, but carries meaningful structure across toolcalls:}
We analyze the types of tools invoked and their execution times in \catraces{}. \autoref{fig:tool_call_dist} shows that tool execution times vary significantly across tool types and exhibit long-tailed behavior. For example, \texttt{grep} has short and tightly clustered durations, whereas \texttt{bash} exhibits longer and more variable execution times. 
This variability translates directly into large fluctuations in the time between successive serving requests from the same session. \autoref{fig:inter_llm_time} shows that tool executions induce substantial variation in inter-arrival times, leading to irregular reuse intervals for \kvcache{} state. Eviction policies that rely solely on past access patterns (e.g., LRU) may evict state that is about to be reused, resulting in unnecessary evictions. 

Despite variations, tool call durations are not arbitrary. \autoref{fig:tool_examples} shows that tool metadata, such as tool type and arguments, captures meaningful structure in execution behavior. For instance, simple file-system operations (e.g., \texttt{ls}, \texttt{grep}) have short durations, whereas more complex operations (e.g., \texttt{pytest}) exhibit significantly longer execution times. This observation exposes opportunities for \kvcache{} optimization by leveraging tool metadata as signals for reuse prediction, which we discuss further in \xref{sec:design}.


\begin{figure}[t]
    \centering
        \begin{subfigure}{0.45\linewidth}
        \centering
        \begin{lstlisting}
{
  "tool": "bash",
  "args": "ls -la",
  "duration_ms": 49
}
{
  "tool": "bash",
  "args": "cd vllm && git log --all -p --grep='SchedulingPolicy' -- '*.py' | head -300",
  "duration_ms": 1143
}
{
  "tool": "bash",
  "args": "uv run --frozen pytest tests/client
  /test_set_roots.py -xvs",
  "duration_ms": 83333
}
    \end{lstlisting}
        \end{subfigure}
        \hfill
        \begin{subfigure}{0.45\linewidth}
        \centering
\begin{lstlisting}
{
  "tool": "Grep",
  "args": {"path": "vllm/vllm", "pattern": "start_load_kv",
  "duration_ms": 183
}
{
  "tool": "Glob",
  "args": "**/scheduler*.py",
  "duration_ms": 147
}
{
  "tool": "WebFetch",
  "args": "https://gofastmcp.com
  /servers/context",
  "duration_ms": 39988
}
        \end{lstlisting}
        \end{subfigure}
        
        \caption{Examples of tool invocations observed in \catraces{}.}
        \label{fig:tool_examples}
\end{figure}

\section{Implications for Efficient LLM Serving}
\label{sec:limitations}

Coding agent workloads exhibit strong temporal locality, where successive requests within a session share and reuse substantial portions of previously computed \kvcache{} state. Existing LLM serving systems manage \kvcache{} allocation and eviction independently at the granularity of individual requests, without accounting for future reuse across successive requests within a session. This mismatch leads to significantly lower \kvcache{} reuse and system efficiency. 

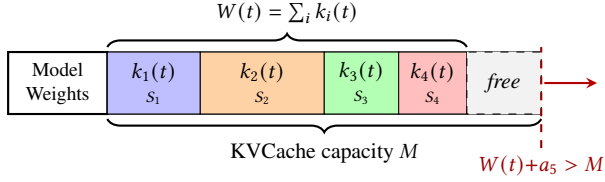
\begin{figure}
\centering
\begin{tikzpicture}[
    scale=0.82, every node/.style={scale=0.82},
    font=\normalsize,
    >=stealth,
    block/.style={draw, minimum height=1.0cm, inner sep=0pt},
    brace/.style={decorate,
                  decoration={brace, amplitude=5pt, mirror},
                  thick},
    label above/.style={above, yshift=4pt},
  ]

  \def\totalW{8.6}   
  \def\weightsW{1.6} 
  \def\cacheW{7.0}   
  \def\barH{1.0}     

  \def\cacheStart{\weightsW}
  \def\cacheEnd{\totalW}

  \def\sOneW{1.5}
  \def\sTwoW{2.0}
  \def\sThreeW{1.2}
  \def\sFourW{1.1}
  \def\freeW{1.2}   

  \draw[block, fill=gray!15, minimum width=\totalW cm]
        (0,0) rectangle (\totalW, \barH);

  \fill[white] (0,0) rectangle (\weightsW, \barH);
  \fill[white]
        (0,0) rectangle (\weightsW, \barH);
  \draw[thick] (0,0) rectangle (\weightsW, \barH);
  \node[align=center, font=\small, fill=white, fill opacity=1,
        text opacity=1, inner sep=1pt]
        at (\weightsW/2, \barH/2)
        {Model\\Weights};

  \def\colors{{"blue!25","orange!35","green!30","red!25"}}
  \def\labels{{"$k_1(t)$","$k_2(t)$","$k_3(t)$","$k_4(t)$"}}
  \def\slabels{{"$S_1$","$S_2$","$S_3$","$S_4$"}}
  \def\widths{{1.5,2.0,1.2,1.1}}

  \fill[fill=blue!25]
        (\cacheStart, 0)
        rectangle (\cacheStart+\sOneW, \barH);
  \draw (\cacheStart, 0) rectangle (\cacheStart+\sOneW, \barH);
  \node[align=center] at (\cacheStart+\sOneW/2, \barH/2)
        {$k_1(t)$\\[-1pt]{\scriptsize$S_1$}};

  \pgfmathsetmacro{\xTwo}{\cacheStart+\sOneW}
  \fill[fill=orange!35]
        (\xTwo, 0) rectangle (\xTwo+\sTwoW, \barH);
  \draw (\xTwo, 0) rectangle (\xTwo+\sTwoW, \barH);
  \node[align=center] at (\xTwo+\sTwoW/2, \barH/2)
        {$k_2(t)$\\[-1pt]{\scriptsize$S_2$}};

  \pgfmathsetmacro{\xThree}{\xTwo+\sTwoW}
  \fill[fill=green!30]
        (\xThree, 0) rectangle (\xThree+\sThreeW, \barH);
  \draw (\xThree, 0) rectangle (\xThree+\sThreeW, \barH);
  \node[align=center] at (\xThree+\sThreeW/2, \barH/2)
        {$k_3(t)$\\[-1pt]{\scriptsize$S_3$}};

  \pgfmathsetmacro{\xFour}{\xThree+\sThreeW}
  \fill[fill=red!25]
        (\xFour, 0) rectangle (\xFour+\sFourW, \barH);
  \draw (\xFour, 0) rectangle (\xFour+\sFourW, \barH);
  \node[align=center] at (\xFour+\sFourW/2, \barH/2)
        {$k_4(t)$\\[-1pt]{\scriptsize$S_4$}};

  \pgfmathsetmacro{\xFree}{\xFour+\sFourW}
  \fill[fill=gray!10]
        (\xFree, 0) rectangle (\totalW, \barH);
  \draw[dashed] (\xFree, 0) rectangle (\totalW, \barH);
  \node at (\xFree+\freeW/2, \barH/2) {\textit{free}};

  \draw[brace] (\cacheStart, -0.08) -- (\totalW, -0.08)
        node[midway, below=6pt] {KVCache capacity $M$};

  \draw[decorate,
        decoration={brace, amplitude=5pt},
        thick]
        (\cacheStart, \barH+0.08) -- (\xFree, \barH+0.08)
        node[midway, above=6pt]
             {$W(t)=\sum_i k_i(t)$};

  \pgfmathsetmacro{\reqStart}{\xFree}
  \pgfmathsetmacro{\reqEnd}{\totalW+1.0}  

  \draw[->, thick, red!70!black]
        (\totalW+0.05, \barH/2) -- (\totalW+0.9, \barH/2);

  \draw[dashed, thick, red!60!black]
        (\totalW, -0.55) -- (\totalW, \barH+0.08);

  \node[red!60!black, below] at (\totalW, -0.55) {$W(t){+}a_5 > M$};

\end{tikzpicture}
\caption{
  XPU memory layout during LLM serving.
  After reserving memory for model weights, the remaining
  capacity $M$ holds {\kvcache} blocks for active sessions
  $\mathcal{S}_t = \{S_1, \ldots, S_4\}$.
  The working set $W(t) = \sum_i k_i(t)$ grows as sessions
  accumulate context.
  When a new request $r_5$ requires $a_5$ additional blocks
  that exceed $M$, the system must evict blocks from an existing session.
}
\label{fig:model}
\end{figure}

We introduce a lightweight formal model to precisely characterize the resource dynamics of {\kvcache} management in LLM serving and highlight the implications for coding agents. Figure~\ref{fig:model} illustrates the memory layout that motivates our notation.

Let $M$ denote the total {\kvcache} block capacity of an inference node (i.e., XPU memory available for {\kvcache} after accounting for model weights and runtime overhead).
At any time $t$, let $\mathcal{S}_t = \{S_1, S_2, \ldots\}$ be the set of \emph{active sessions} co-located on the node.
Each session $S_i$ has a contiguous, monotonically growing prefix of {\kvcache} blocks $d_i$, out of which $k_i(t)$ are resident in XPU memory at time $t$. The \emph{active {\kvcache} working set} is $W(t) = \sum_{i \,\in\, \mathcal{S}_t} d_i$, and the system operates within memory budget (requires no eviction) as long as $W(t) \le M$.

When session $S_i$ issues a new request $r_i$ at time $t$, it requires $d_i$ {\kvcache} blocks in total to serve the full context.
Of these, $k_i(t))$ blocks already reside in memory (the \emph{reusable prefix overlap}), so the number of \emph{additional} blocks that must be allocated is $a_i(t) = d_i - k_i(t)$.
If $W(t) + a_i(t) \le M$, the request is admitted immediately. Otherwise, the system must \emph{evict} $a_i(t)$ blocks from one or more other sessions before admission.

\vspace{0.05in}
\noindent\textbf{Implication \#1: FCFS scheduling thrashes \kvcache{} of long-running agent sessions:}
At any given time $t$, a session $S_j$ can be active (i.e., executing a request on the {\xpu}), queued (i.e., request is waiting to be executed due to insufficient {\xpu} memory), or inactive (i.e., waiting upon user or the environment to submit the next request). Existing systems which admit requests in FCFS order interleave requests from multiple queued sessions, which attempts to increases $k_{i}$ for many different sessions. As the working set grows, it competes for limited \xpu{} memory and can exceed the available HBM capacity. When this occurs, \kvcache{} state from one session is evicted to accommodate requests from other sessions. 
Since this evicted state is often reused by subsequent requests within the same session, FCFS scheduling leads to repeated eviction and recomputation (or movement) of state, resulting in \kvcache{} thrashing.

\vspace{0.05in}
\noindent\textbf{Implication \#2: LRU eviction does not account for future \kvcache{} reuse when toolcalls return:} When $W(t) + a_i > M$, the system faces a trade-off: admitting $r_i$ requires evicting blocks whose $\tau_j$ may be small, incurring recomputation overhead and reducing system
goodput. However, not all evictions are equally costly. Let $\tau_i$ denote the \emph{time to next reuse} of session $S_i$'s resident blocks, i.e., the time until $S_i$ issues its next request and those blocks are accessed again. Evicting blocks from a session with small $\tau_i$ forces an imminent recomputation (or data movement) penalty; evicting from a session with large $\tau_i$ is comparatively cheap. An ideal eviction policy therefore selects $j^* = \operatorname*{arg\,max}_{j \,\in\, \mathcal{S}_t,\; j \ne i} \tau_j$, retaining sessions whose {\kvcache} will be needed soonest.  

In-practice, when serving \codingagent{}s with multiple long-running, \kvcache{} working set is likely to exceed the available \xpu{} memory. Since $\tau_i$ is unknown, a workload-agnostic policy (e.g., LRU) resolves this tension using only past access recency, which can lead to priority inversions and blocks have high reuse value are evicted and subsequent requests from the same session require these blocks to be recomputed or moved back into memory. Repeated eviction and restoration of state leads to high overhead in recomputation and data movement.


\section{\system{} Design}
\label{sec:design}

\begin{figure*}[t]
  \centering
  \includegraphics[width=1\linewidth]{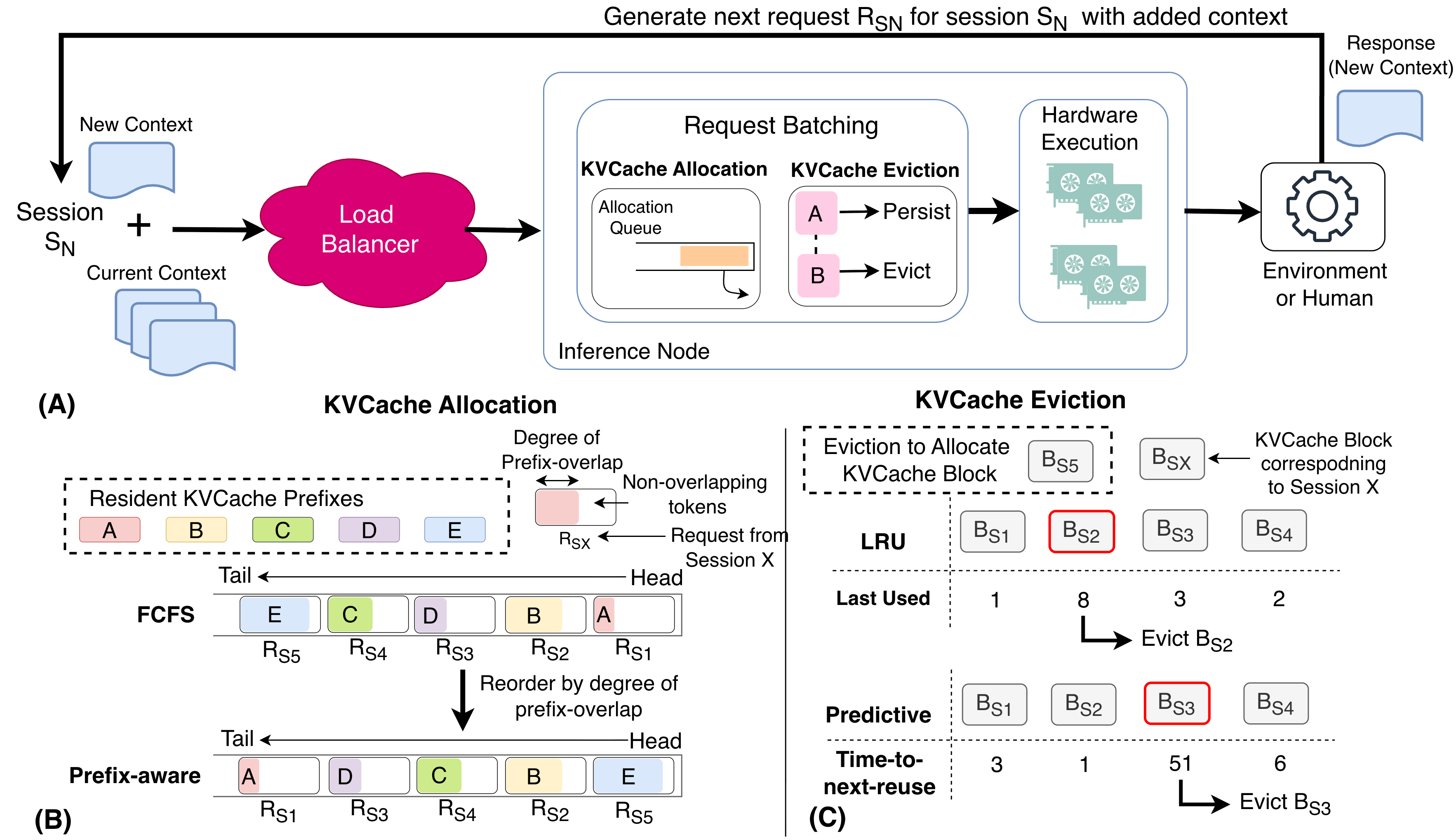}
  \caption{(A) End-to-end workflow of a session through a typical LLM serving system. (B) Prefix-aware request scheduling prioritizes requests by degree of overlap with {\kvcache} resident on the {\xpu} memory, leading to higher {\kvcache} reuse. (C) Demonstration of predictive {\kvcache} eviction choosing the block with the highest predicted reuse probability, rather than purely recency-based heuristics like LRU.}\label{fig:design-overview}
\end{figure*}

\begin{table*}[t!]
\centering
\footnotesize
\setlength{\tabcolsep}{5pt}
\begin{tabular}{@{}p{0.11\textwidth}p{0.26\textwidth}p{0.26\textwidth}p{0.26\textwidth}@{}}
\toprule
\textbf{Property} & \textbf{Workload Characteristic} & \textbf{Baseline Limitation} & \textbf{{\projectname} Design Choice} \\
\midrule
Closed-loop sessions & User tasks trigger multi-turn sessions of LLM calls and tool executions. &
Per-request metrics (TTFT, TBT) do not capture session-level efficiency. &
Optimize for end-to-end session completion for closed-loop workloads. \\[2pt]
Expanding \kvcache{} prefixes &
Sessions accumulate large {\kvcache} prefixes that compete for limited {\xpu} memory. &
FCFS request scheduling disregards {\kvcache} residency which leads to thrashing across sessions. &
Prefix-aware scheduling forms batches in priority order to maximize {\kvcache} reuse. \\[2pt]
Variable inter-request times &
Tool execution durations vary by orders of magnitude due to different tool types and arguments. &
LRU eviction uses only access recency that which leads to priority inversions w.r.t. future re-use. &
Predictive eviction estimates the order of future reuse from tool metadata to guide eviction decisions. \\
\bottomrule
\end{tabular}
\caption{Summary of key findings. Each row identifies a characteristic of \codingagent{} workloads, the limitation it exposes in existing systems, and the corresponding {\projectname} design choice.}
\label{tab:findings}
\end{table*}

Drawing from our measurements and analysis, we present {\projectname} (Table~\ref{tab:findings}). Unlike existing systems that optimize {\kvcache} management for individual request processing commonly found in chatbots, {\projectname} maximizes {\kvcache} reuse across requests in long-running, closed-loop agent sessions (e.g., \codingagent{} workloads). Higher {\kvcache} reuse reduces the eviction overhead (recomputation and data movement overhead of {\kvcache} across storage tiers) and allows executing more sessions concurrently which improves the end-to-end session performance and the system's token goodput. 

\vspace{0.05in}
\noindent\textbf{Workflow:} Figure~\ref{fig:design-overview} (A) illustrates the end-to-end workflow of a session in an LLM serving system with {\projectname} integrated at the inference node. Incoming requests from active sessions are routed by the load balancer to an inference node. Since existing load balancers typically use prefix-match based routing~\cite{vllm_prefixaware_routing, bentoml_prefix_aware_routing, linkedin_prefix_aware_routing}, i.e., a request is routed to the inference node with largest reusable {\kvcache} prefix resident in the {\xpu} memory, requests from the same session, which share the growing {\kvcache} prefix, are likely to be routed to the same inference node. {\projectname} optimizes {\kvcache} management at each inference node using two key techniques: (a) prefix-aware request scheduling, and (b) predictive {\kvcache} eviction.

\subsection{Prefix-aware Request Scheduling}
{\projectname} leverages a prefix-aware request scheduler that prioritizes requests that can reuse {\kvcache} already resident in {\xpu} memory. Specifically, at time $t$, {\projectname} selects request $r_i$ which requires fewest additional blocks $a_i(t)$ to be allocated as the next request to dispatch.

Although greedy prefix-aware scheduling has been explored in prior systems such as SGLang, we observe that it is particularly effective for \codingagent{} workloads because it directly minimizes the \emph{thrashing} effect on other sessions which will eventually re-use their evicted blocks as well. Furthermore, dispatching requests in order of $a_i(t)$ approximates \emph{shortest-job-first}~\cite{lwl} scheduling which additionally minimizes overall system queueing at the expense of per-request responsiveness, since for closed-loop agent workloads, optimizing end-to-end session completions is more important than minimizing per-request TTFT or TBT, as no user is waiting on any individual request.

\subsection{Predictive {\kvcache} Eviction}

While prefix-aware scheduling reduces the overhead of evictions, they remain inevitable because sessions keep accumulating context across turns and overall memory required increases. Ideally, the optimal eviction strategy follows Belady's rule~\cite{belady}, i.e., at time $t$, evict blocks whose next access time $\tau_i(t)$ is furthest into the future, so that the evicted memory can be used for other sessions for the entire duration $\tau_i(t)$. Conversely, if the evicted blocks will be required almost immediately, and because $W(t) > M$, to allocate them again, blocks from some other session must be evicted at $\tau_i(t)$, which increases the overall eviction overhead incurred.

{\projectname} proposes a predictive {\kvcache} eviction policy that attempts to approximate the optimal eviction strategy by evicting blocks in decreasing order of $\tau_i(t)$. Our key insight is that although precisely predicting future {\kvcache} reuse time for each session is challenging due to the variability in tool execution times, {\projectname} only needs to predict the relative order of $\tau(t)$ across sessions to identify which blocks to evict. Furthermore, in-practice, we find that accurately identifying the session with the highest $\tau_i(t)$ is sufficient when using {\projectname}, and lightweight estimators based on function names and arguments for each tool call can provide sufficient accuracy.

\vspace{0.05in}
\noindent\textbf{Predictor:} For a session $S_{i}$ that generated a tool call with metadata $m = (\text{tool\_name}, \text{tool\_args})$ at time $T_{i}$, our predictor estimates it's next expected use time $\mathbb{E}[\tau_i(t)]$ using historical information, and uses these estimates to calculate the relative ordering to make its eviction decisions. Specifically, at time $t$, {\projectname} estimates the expected remaining time for the tool call to complete, i.e., $\mathbb{E}[\tau_i(t) \mid \tau_i(t) - T_{i} > t - T_{i}]$ by analyzing distributions of tool execution durations with similar metadata $m$ from historical sessions with similar characteristics e.g., same user, same project.

\begin{figure}
    \centering
    \includegraphics[width=1\linewidth]{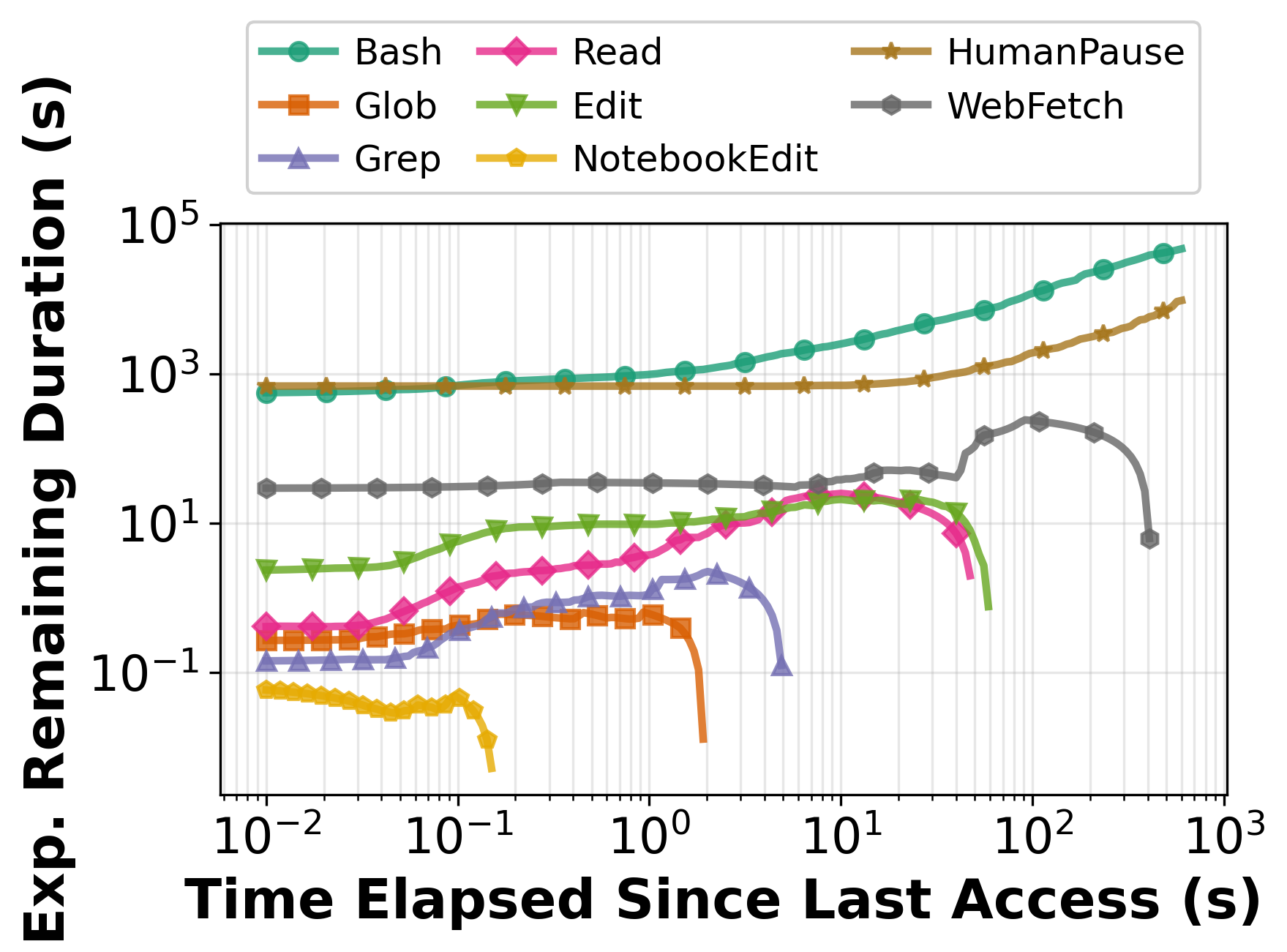}
    \caption{Distribution of tool execution times with respect to time elapsed since last access as observed in \catraces{}.}
    \label{fig:design-tool-aging}
\end{figure}

\vspace{0.05in}
\noindent\textbf{Predictor Training:} Figure~\ref{fig:design-tool-aging} shows that relying solely on execution duration distributions of tools with the same names is insufficient to make these estimates accurately because execution times may vary significantly based on the tool arguments and the environment; for instance, \texttt{Bash} calls can have highly variable execution durations depending upon the arguments (e.g., Table~\ref{tab:bash-clusters}) or the coding repository context (e.g., running \texttt{pytest} to run tests depends upon the specific test suite). To capture the effect of tool call arguments, {\projectname} semantically clusters\footnote{We use \texttt{KMeans}~\cite{kmeans} to generate clusters. We upper bound
the number of clusters to a large value to achieve maximum granularity,
allowing the model to distinguish fine-grained argument patterns within
each tool type.} historical samples based on the TF-IDF embeddings~\cite{tfid_vector, tfidf_doc} of their arguments, and uses these per-cluster distributions instead of per-tool distributions in such cases. TF-IDF encodes tool arguments into vectors by assigning higher weight to terms that are frequent within an argument but rare across other arguments \footnote{Each tool argument is mapped to a TF-IDF vector of a maximum size of 5000 lexicons. We use scikit-learn's implementation of $\ell_2$ norm~\cite{tfidf_doc} to assign weights.}.

\begin{table}
\centering
\small
\setlength{\tabcolsep}{3pt}
\begin{tabular}{@{}lrrrr@{}}
\toprule
Command Pattern & P50 & P90 & P99 & n \\
\midrule
Python one-liners (\texttt{python -c}) & 0.1s & 0.2s & 0.4s & 9 \\
Type checking (\texttt{mypy}) & 10s & 56s & 182s & 46 \\
Docker builds (\texttt{docker compose}) & 22s & 89s & 152s & 30 \\
Git operations (\texttt{git add / status}) & 0.1s & 34s & 97s & 34 \\
\bottomrule
\end{tabular}
\caption{Bash tool call execution time distributions clustered by argument types. Columns P50, P90, and P99 show percentiles of tool execution times. $n$ is the number of datapoints in the cluster.}
\label{tab:bash-clusters}
\end{table}


\subsection{Implementation}

We have implemented {\projectname} using vLLM~\cite{vllm} in \textasciitilde2,500 lines of Python code and extend the batch scheduler and {\kvcache} manager to support {\projectname}'s predictive block eviction mechanisms. Specifically, to implement {\projectname}'s policies atop the vLLMs existing batch scheduling and {\kvcache} management layer, we extend their {\kvcache} block manager to associate additional session-level metadata with each {\kvcache} block, i.e., \text{tool\_name}, \text{tool\_args} and $T_{i}$ (time when the tool call was generated), which can be parsed by the vLLM engine itself. 

When a request completes and its blocks become unreferenced (i.e., the reference count drops to zero), those blocks retain their metadata and are inserted into an eviction heap for ordered eviction. \projectname{} replaces the default LRU eviction policy: when unreferenced blocks are added, our lightweight predictor quickly estimates the $\mathbb{E}[\tau_i(t)]$ based on the block's attached $S_{i}$, and adds the block to the eviction heap with the prediction as its priority value. Note that \projectname{} maintains predictions across blocks that share the same session metadata and multiple \kvcache{} blocks belonging to the same session do not require invoking the predictor independently for each one.

However, as tool call executions progress, the predicted $\mathbb{E}[\tau_i(t)]$ value for a \kvcache{} block becomes stale; as time elapses since the last access, the expected time-to-next-reuse shifts, and the stored prediction no longer reflects the current reuse likelihood of the block. To account for this, \projectname{} periodically re-evaluates all unreferenced blocks through the predictor and rebuilds the eviction heap with updated $\mathbb{E}[\tau_i(t)]$ estimates. The rebuild frequency is controlled by a tunable parameter $N_{\text{rebuild}}$ (in engine iterations): smaller values yield fresher estimates at higher CPU scheduling overhead, while larger values reduce overhead at the cost of stale predictions. In our evaluation, we set $N_{\text{rebuild}} = 3$ based on empirical observation that this balances prediction freshness and overhead.


\section{Evaluation}\label{sec:evaluation}
We evaluate {\projectname} using real multi-turn coding-agent traces sampled from the \catraces{} dataset. Our evaluation seeks to answer the following questions:

\squishlist
\item How much {\projectname} improve end-to-end performance compared to baselines?
\item How much  prefix-aware scheduling and predictive {\kvcache} eviction contribute to end-to-end performance?
\item Does {\projectname} reduce {\kvcache} data movement?
\item Does {\projectname} introduce non-negligible scheduling overheads?
\squishend

\subsection{Experimental Setup}

\vspace{0.05in}
\noindent\textbf{Testbed:} We ran experiments on a server with 2$\times$ H200 \gpu{}s and AMD EPYC 9534 64-core CPU with Tensor Parallelism~\cite{megatron} (TP=2) enabled, with a combined GPU memory of 282GB. All experiments run with chunked-prefill enabled and we use the default value of max. 512 tokens per chunk~\cite{default_chunked_prefill}. 

\vspace{0.05in}
\noindent\textbf{Model:} Similar to prior works~\cite{sarathi-serve, distserve}, we use a $32B$ parameter model, Qwen2.5-Coder-32B-Instruct~\cite{qwen}, for our evaluation. Our prototype can execute any model, including models with grouped-query attention~\cite{gqa} or multi-query attention~\cite{mqa}, without any modifications to model weights. 

\vspace{0.05in}
\noindent\textbf{Workloads:} We replay the traces collected in our dataset for our experiments deterministically. We randomly sample $80\%$ of the sessions for \emph{offline} training of the predictor, and use the remaining traces for our experiments. We assume that the system deploying \projectname{} routinely logs request traces which can be used as the training data.
In practice, trace distributions may drift over time as coding repositories and tool usage patterns evolve; in such cases, \projectname{}'s estimators can be periodically retrained on recent traces to adapt to distribution shift, though we leave a thorough study of drift robustness to future work.

In our traces, users interactively respond to agent output to send a new request with the same {\kvcache} prefix. Users can have long idle periods. During replay, we consider human idle periods as session boundaries. Our evaluation includes resumed sessions after human pause and these sessions have a significantly larger {\kvcache} to prefill due to previous history. This is acceptable because it mimics our scheduler's decisions: we always evict {\kvcache} from idle sessions before evicting from sessions with toolcall induced delays.

\vspace{0.05in}
\noindent\textbf{Methodology:} Our experiments launch $N$ independent \codingagent{} sessions concurrently on a single serving instance and runs until all sessions complete. We evaluate performance under increasing load. We repeat each experiment several times to ensure stability across runs.

\begin{figure}[t]
    \centering
    \includegraphics[width=1\linewidth]{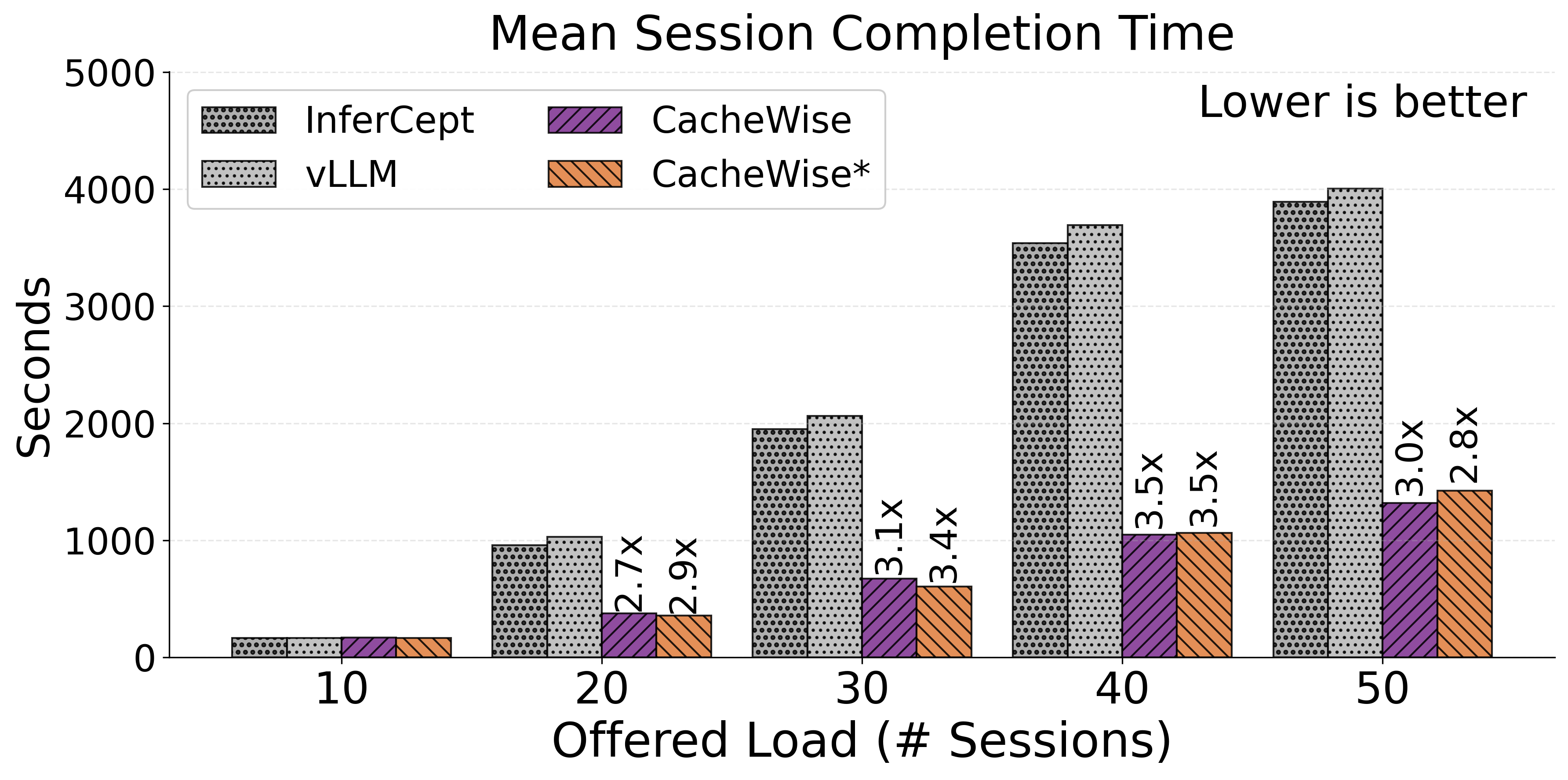}
    \caption{Session completion time (in seconds) for different {\kvcache} management systems under the \codingagent{} workload (sampled from \catraces{}).}
    \label{fig:eval-makespan}
\end{figure}

\vspace{0.05in}
\noindent\textbf{Baselines:}
We compare {\projectname} against: (a) \textbf{vLLM}~\cite{vllm} with block-level LRU {\kvcache} eviction and FCFS admission, (b) \textbf{InferCept}~\cite{infercept} with FCFS scheduling and moving-average-based eviction that predicts tool call durations from recent history. 
Although, we did not directly evaluate against SGLang, which already implements prefix-aware scheduling, our ablations (described below) cover this scenario as well.

\begin{figure*}[t]
    \centering
    \includegraphics[width=1\linewidth]{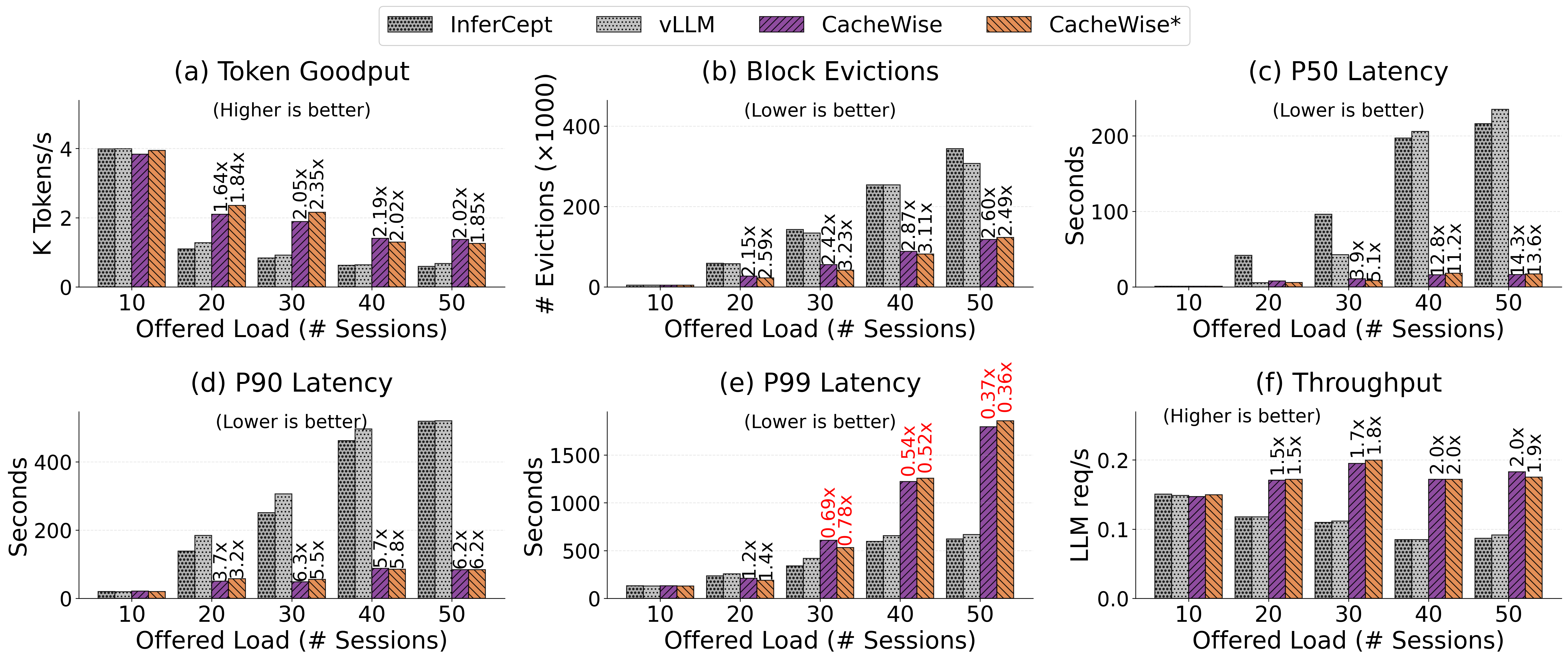}
    \caption{Serving efficiency of different {\kvcache} management systems under the \codingagent{} workload. (a) Goodput in-terms of the number of useful tokens generated per-unit time. (b)-(e) LLM request latency distribution (in seconds). (f) LLM request throughput (in number of requests completed per second).}
    \label{fig:eval-serving-efficiency} 
\end{figure*}

\vspace{0.05in}
\noindent\textbf{Ablations:} We also present (a) {\projectname}*, a version of {\projectname} that has access to ground truth tool latencies to make evictions (b) \textbf{vLLM(Prefix-aware)} and (c) \textbf{InferCept(Prefix-aware)} which add prefix-aware scheduling to these baselines.

\vspace{0.05in}
\noindent\textbf{Metrics:} We compare (1) token goodput (rate of new tokens computed per unit time), request throughput (LLM requests/second) (2) request completion time, (3) session completion time (the sum of LLM request latencies across a single session without the tool execution durations. We exclude tool execution durations as they are determined by the external environment and independent of the serving system), and (4) {\kvcache} transfer volume (the amount of data moved due to evictions).

\subsection{End-to-End Performance Improvements}

\vspace{0.05in}
\noindent\textbf{Session-level performance:} Figure~\ref{fig:eval-makespan} shows session-level performance across varying load levels. Observe that at lower load levels ($N\leq10$), all systems exhibit similar performance, which is expected as there is negligible memory ({\kvcache}) contention at low load levels, leaving sufficient \gpu{} memory capacity for all sessions to execute concurrently. At higher load levels ($N>10$), {\projectname} consistently outperforms the baselines, achieving 2.7$\times$--3.5$\times$ lower completion time as compared to vLLM and InferCept. Further, {\projectname} achieves performance comparable to the {\projectname}* with ground-truth tool latencies, indicating that our lightweight estimators offer sufficient accuracy to identify which sessions blocks should be evicted. In some configurations, \projectname{} marginally outperforms 
\projectname{}*, which we attribute to scheduling dynamics induced 
by differences in eviction decisions; evicting different blocks 
changes which requests can be scheduled with prefix-aware scheduling, 
which in turn affects the order and timing of requests being processed.

Figure~\ref{fig:eval-serving-efficiency} shows our evaluation of token goodput, {\kvcache} block evictions, request-level latency, and throughput with respect to system load.
\vspace{0.05in}

\noindent\textbf{Token goodput:} Observe that for $N>10$, {\projectname} consistently outperforms vLLM and InferCept with respect to token goodput (1.64x-2x improvement). {\projectname} maximizes {\kvcache} reuse, and consistently reduces the number of evicted {\kvcache} blocks by 2-2.6x compared to vLLM and InferCept, which enables it to increase the useful tokens generated per-unit time. Also observe that the token goodput decreases with increasing load for all systems. This is expected, because at higher load more sessions compete for \gpu{} memory, causing more {\kvcache} evictions and recomputations for all systems. However, {\projectname} minimizes the number of unnecessary evictions and recomputations, even at higher loads.

\vspace{0.05in}
\noindent\textbf{Request throughput:} At higher load levels ($N>10$), {\projectname} consistently out-performs vLLM and InferCept. {\projectname} achieves 1.5$\times$--2$\times$ higher throughput with respect to baselines. This is expected since higher token goodput is directly correlated with higher rate of request completion. 

\vspace{0.05in}
\noindent\textbf{Request latency:}
{\projectname} outperforms both vLLM and InferCept in terms of median (P50), and P90 request completion times across all load levels. At P50, {\projectname} achieves a 13--14$\times$ lower latencies as compared to the baselines, underscoring the benefit of {\projectname}’s better resource management. We observe that {\projectname} exhibits higher P99 request latencies, especially at higher load levels. This is due to {\projectname}’s prefix-aware scheduling policy that defers new requests that lack history and is acceptable---session completion time is the appropriate user-facing metric, rather than request tail latency.

\subsection{Performance Ablations}\label{sec:perf_breakdown}

\begin{figure}[t]
  \centering
  \begin{subfigure}{\columnwidth}
    \centering
    \includegraphics[width=\linewidth]{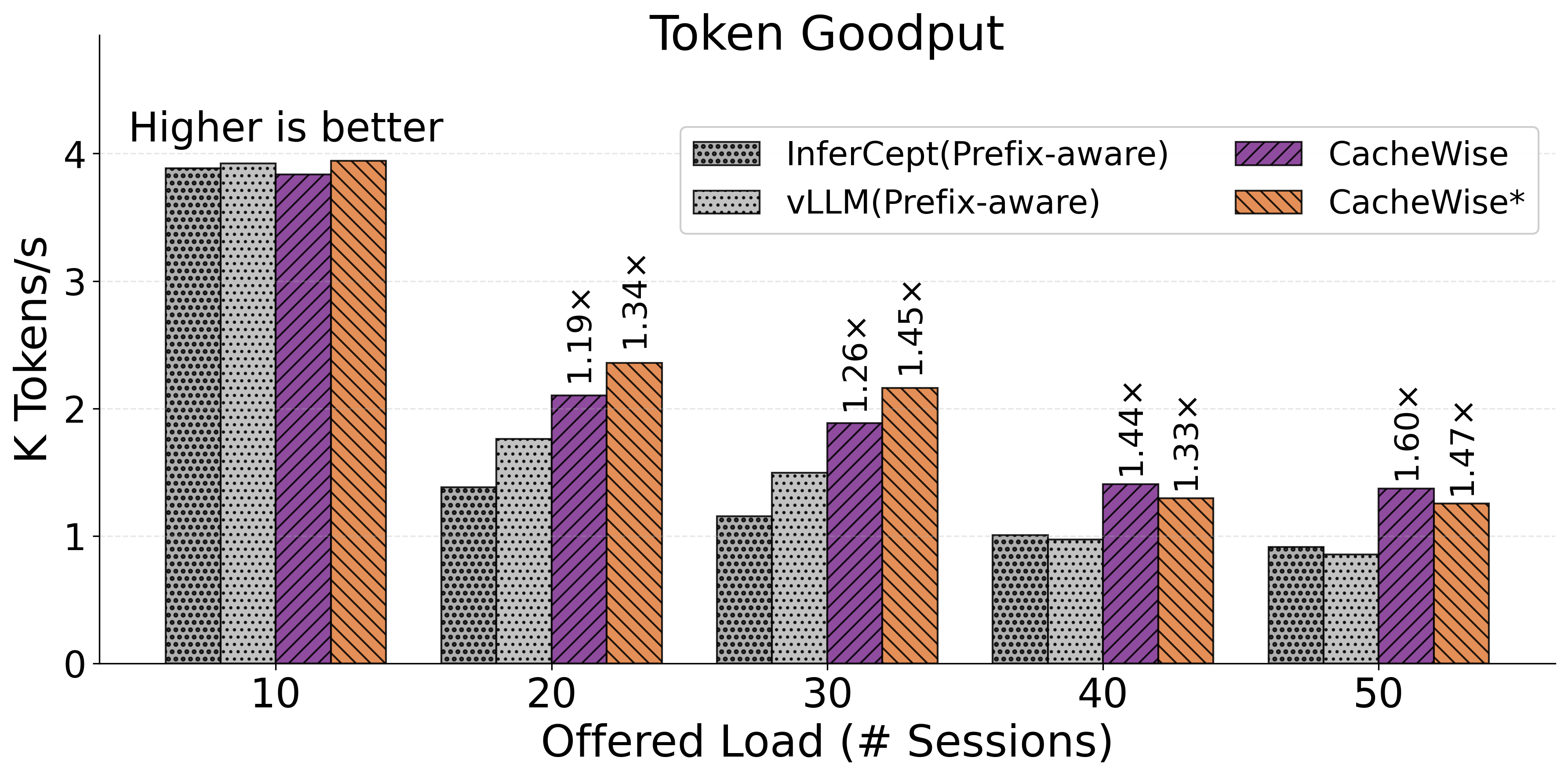}
    \caption{Token goodput.}
    \label{fig:predictive-ablation-goodput}
  \end{subfigure}
  \begin{subfigure}{\columnwidth}
    \centering
    \includegraphics[width=\linewidth]{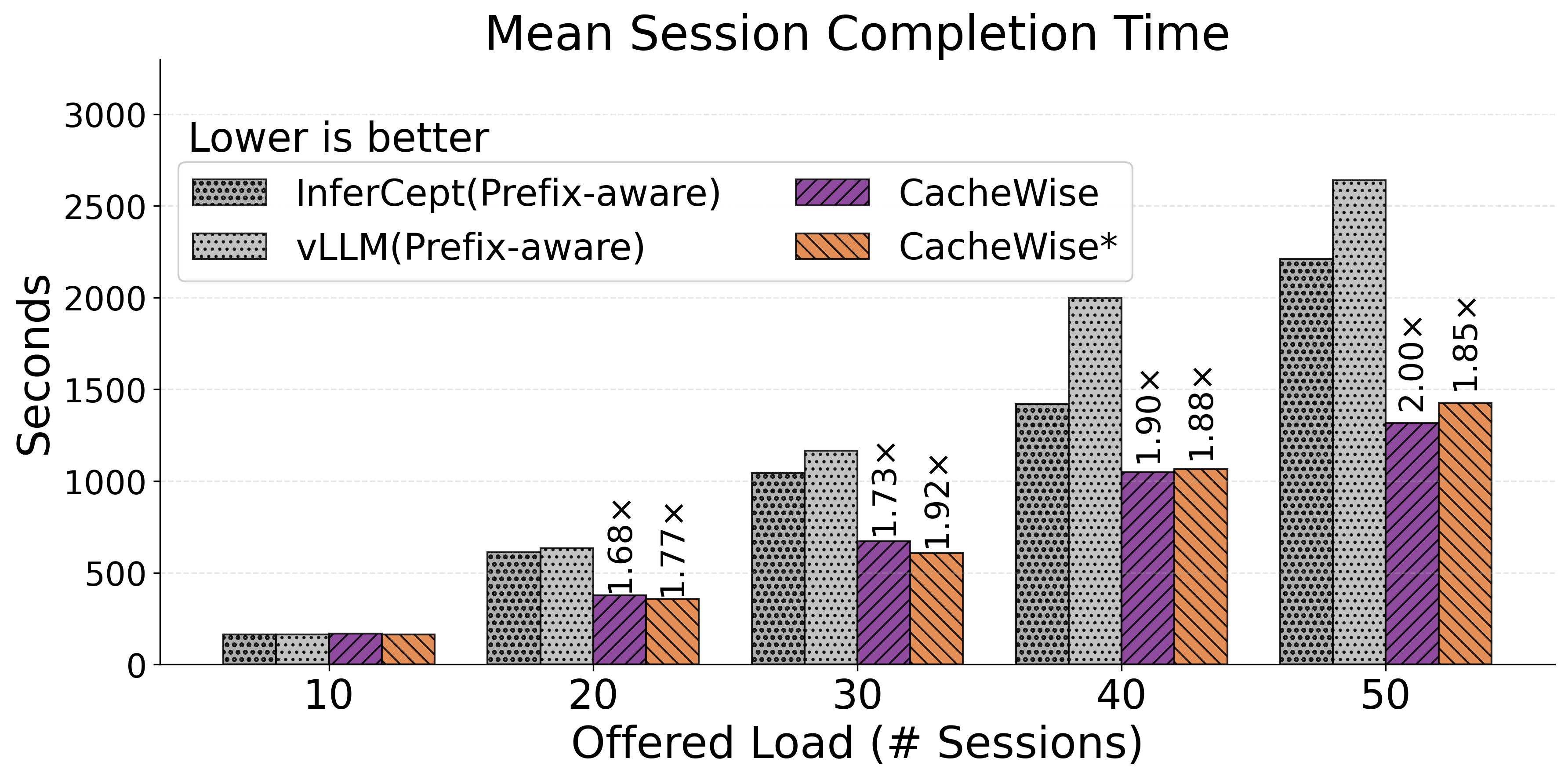}
    \caption{Mean session completion time.}
    \label{fig:predictive-ablation-session}
  \end{subfigure}
  \caption{Impact of {\projectname}'s predictive {\kvcache} eviction. vLLM(Prefix-aware), InferCept(Prefix-aware), \projectname{}, and \projectname{}* with prefix-aware scheduling enabled to isolate impact of {\kvcache} eviction while keeping the scheduling policy fixed.}
  \label{fig:eval-lookahead-ablation}
\end{figure}

To better understand the impact of different optimizations in {\projectname}, we compare it against vLLM(Prefix-aware) and InferCept(Prefix-aware) which use the same prefix-aware scheduling as in {\projectname} but do not have access to the predictive {\kvcache} eviction policy.

\vspace{0.05in}
\noindent\textbf{Impact of predictive {\kvcache} management:} Figure~\ref{fig:eval-lookahead-ablation} shows the gains from predictive {\kvcache} eviction, where {\projectname} outperforms both respect to token goodput and session completion time, achieving 1.2$\times$-1.6$\times$ higher token goodput than baselines. Consequently, {\projectname} also achieves \textasciitilde1.7$\times$-2$\times$ lower session completion time than vLLM(Prefix-aware) and InferCept(Prefix-aware). 

InferCept(Prefix-aware) computes tool execution durations by taking moving average of the prior tool executions in the session, disregarding the tool name, arguments, and the last accessed time of the \kvcache{} blocks, leading to inaccurate time-to-next-reuse decisions, and hence suboptimal eviction decisions.

\begin{figure}[t]
  \centering
  \begin{subfigure}{\columnwidth}
    \centering
    \includegraphics[width=\linewidth]{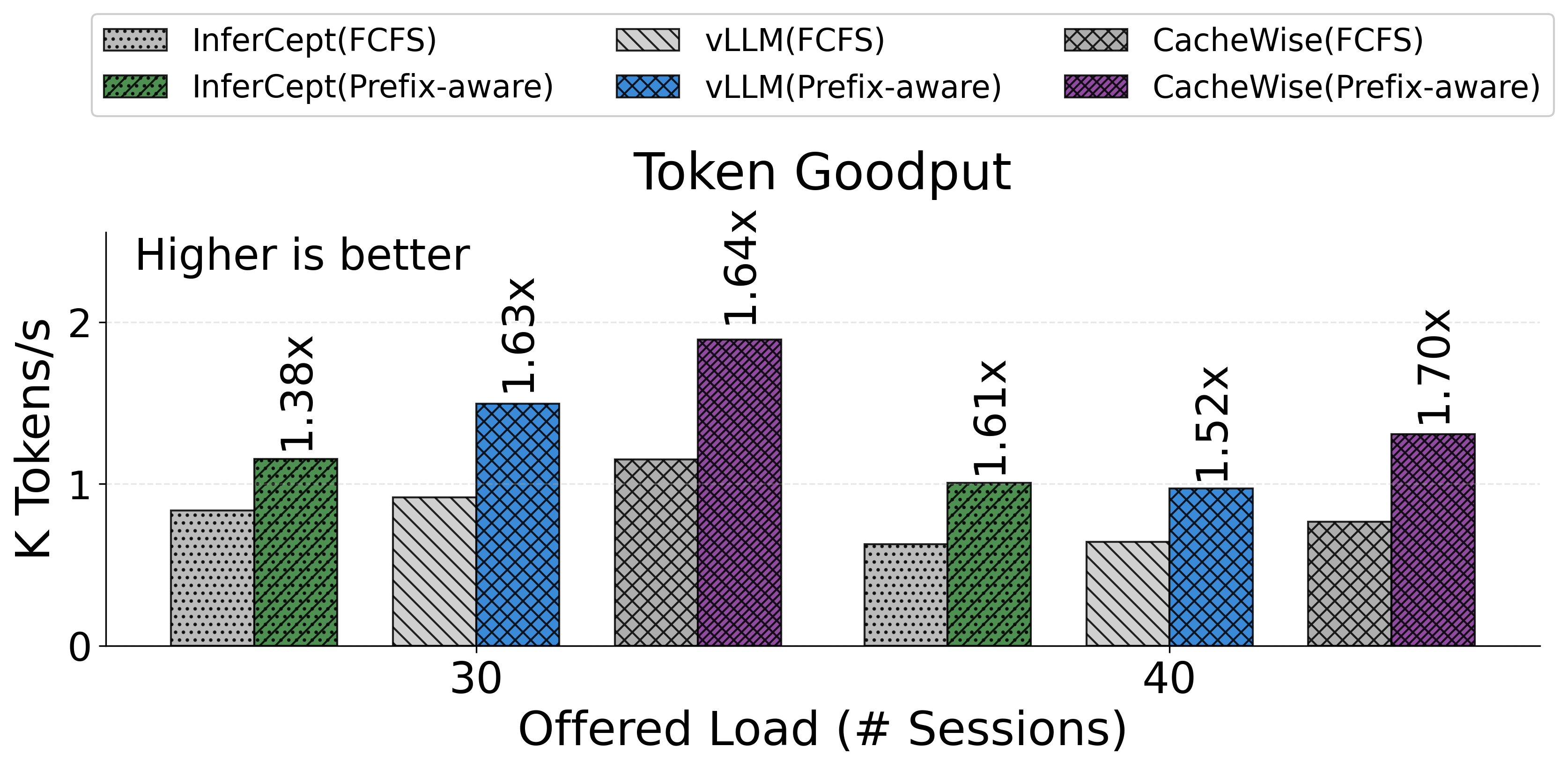}
    \caption{Token goodput.}
    \label{fig:sched-ablation-goodput}
  \end{subfigure}
  \begin{subfigure}{\columnwidth}
    \centering
    \includegraphics[width=\linewidth]{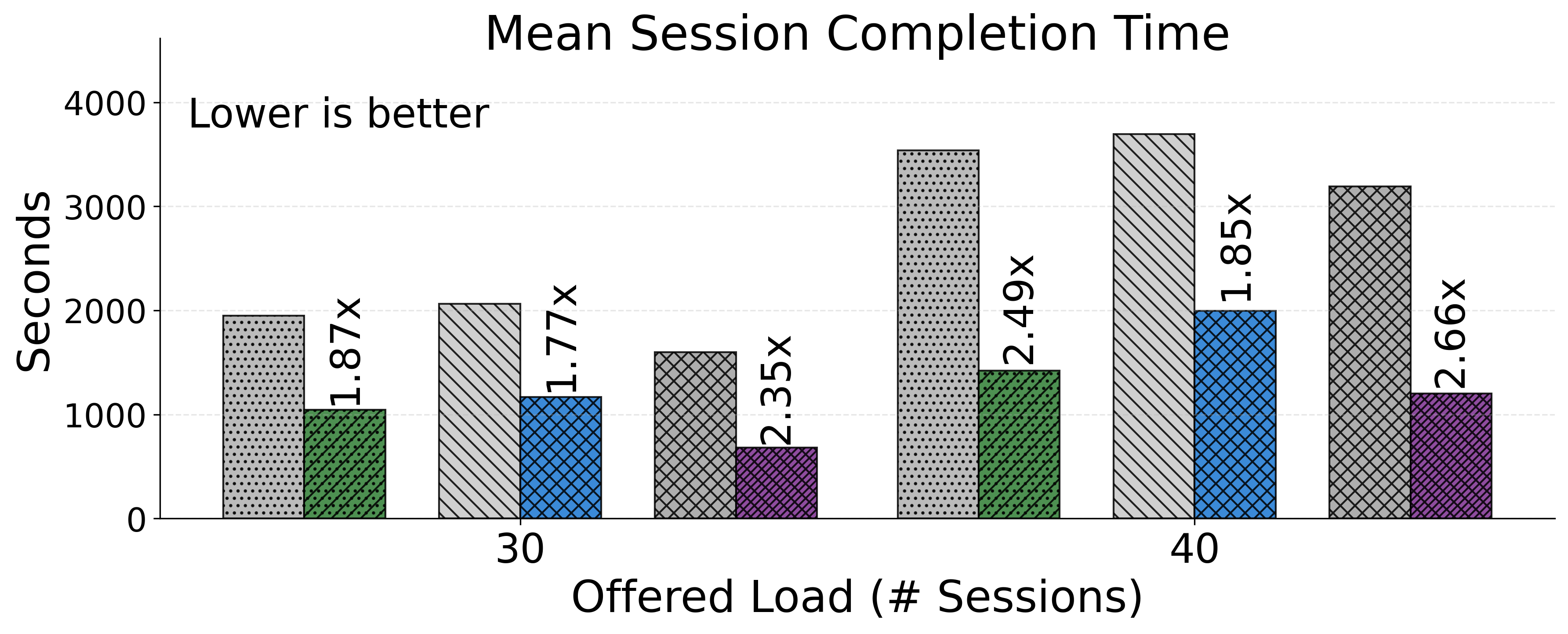}
    \caption{Mean session completion time.}
    \label{fig:sched-ablation-session}
  \end{subfigure}
    \caption{Impact of prefix-aware scheduling policy. For each system, we compare the FCFS variant with the corresponding Prefix-aware variant.}
    \label{fig:eval-sched-ablation}
\end{figure}

\vspace{0.05in}

\noindent\textbf{Impact of prefix-aware scheduling:} 
Figure~\ref{fig:eval-sched-ablation} shows the effect of prefix-aware scheduling applied independently to all baselines. Prefix-aware scheduling alone improves token goodput by \textasciitilde1.38$\times$--1.64$\times$ at $N=30$ and \textasciitilde1.6$\times$--1.7$\times$ at $N=40$ (Figure~\ref{fig:eval-sched-ablation}(a)), and reduces session completion time by \textasciitilde1.8$\times$--2.35$\times$ at $N=30$ and \textasciitilde1.85$\times$--2.66$\times$ at $N=40$ (Figure~\ref{fig:eval-sched-ablation}(b)). These gains hold even in the absence of accurate \kvcache{} reuse predictions, reinforcing that the two optimizations are independent.

\subsection{{\kvcache} Movement}
Modern LLM serving systems commonly offload evicted {\kvcache} blocks to CPU memory or disk~\cite{mooncake, lmcache} to avoid recomputation. For such systems, {\projectname} reduces the frequency of evictions by retaining reusable blocks rather than repeatedly evicting and recomputing them. These transfers consume significant interconnect bandwidths and can stall \gpu{} execution while data is moved, reducing \gpu{} utilization and token goodput. 

\begin{figure}[h]
    \centering
    \includegraphics[width=1\linewidth]{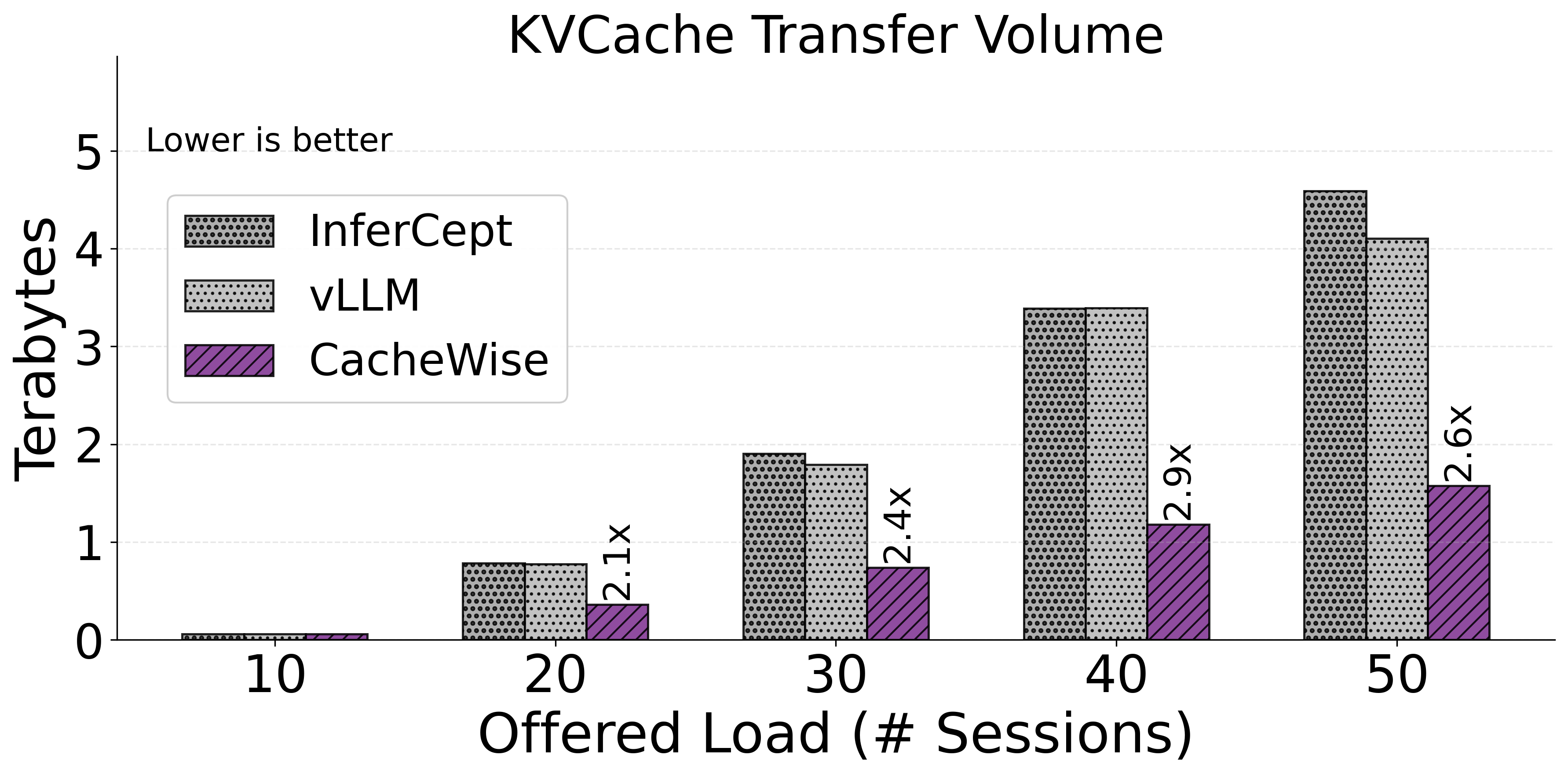}
    \caption{{\kvcache} movement incurred (in TB of data) during the duration of each experiment, by translating eviction count to data size using {\kvcache} block size.}
    \label{fig:kvcache-transfer-vol}
\end{figure}

\begin{figure}[h]
    \centering
    \includegraphics[width=1\linewidth]{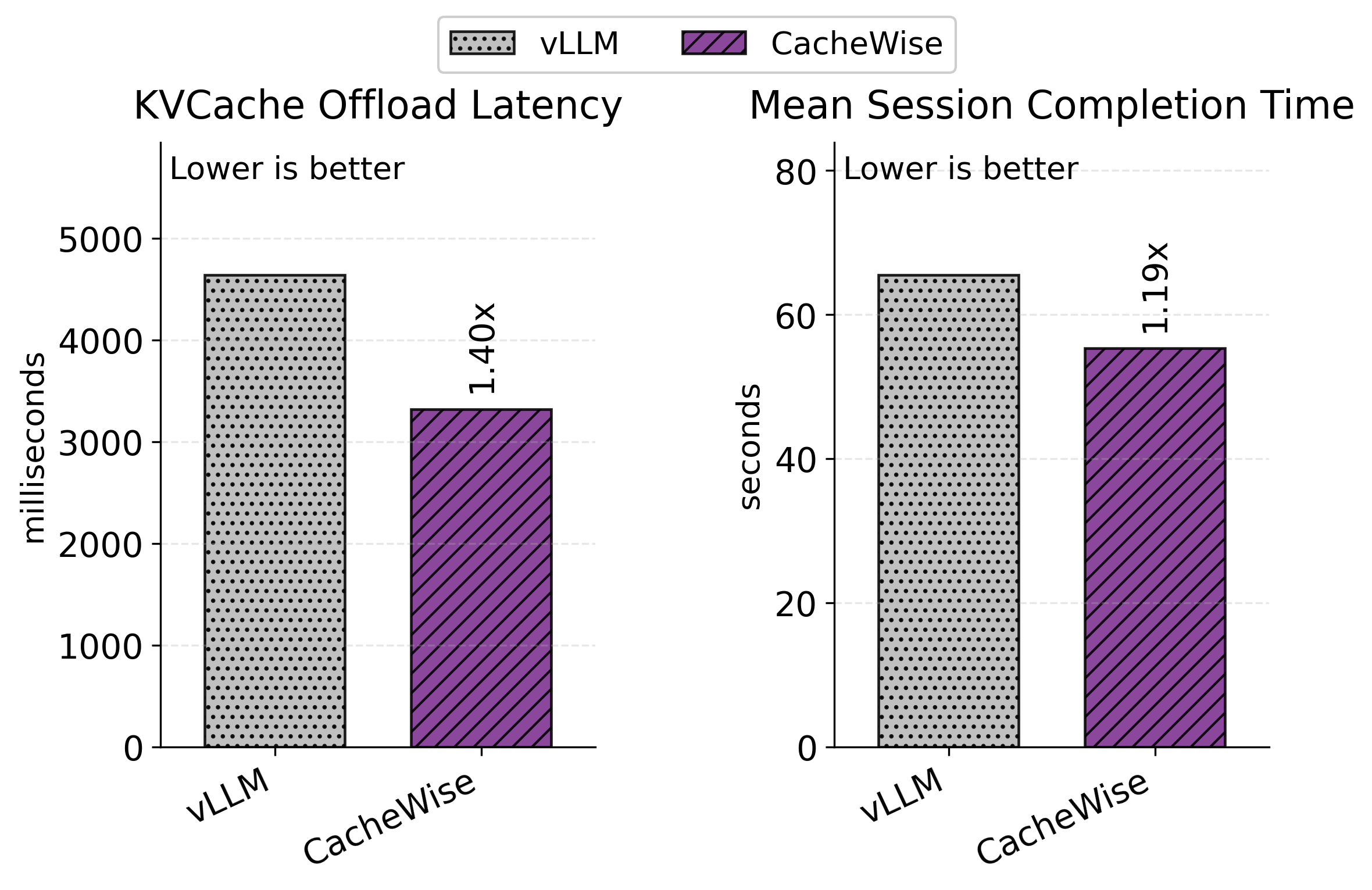}
    \caption{Overhead of {\kvcache} offloading incurred due to transfer of {\kvcache} from GPU--CPU and CPU--GPU over PCIe. The experiment was run with $N$=30 \codingagent{} sessions sampled from \catraces{}. The \kvcache{} offload latency (top-left) is the sum of swap-in and swap-out latencies observed by all the requests in the experiment.}
    \label{fig:kvcache-offload}
\end{figure}

To quantify this effect, we translate {\kvcache} evictions into the corresponding volume of {\kvcache} data transferred and report the cumulative transfer volume over the course of each experiment. Figure~\ref{fig:kvcache-transfer-vol} shows the {\kvcache} data transfer volume from our evaluation. {\projectname} reduces {\kvcache} transfer volume by \textasciitilde2$\times$-2.6$\times$ compared to baselines. 

We further evaluate \projectname{} on a system that physically moves
\kvcache{} blocks rather than recomputing them. In this configuration, evicted blocks are transferred from \gpu{} memory to CPU DRAM and swapped back on demand when the corresponding session resumes, replacing recomputation overhead with idle times due to PCIe transfer latency.
Figure~\ref{fig:kvcache-offload} shows the \kvcache{} offloading latencies and session completion time for $N{=}30$ concurrent sessions. \projectname{} achieves \textasciitilde1.19x lower session completion times than vLLM. The absolute gains under offloading are smaller than under recomputation. This is expected: PCIe transfers are substantially faster than prefill recomputation for large \kvcache{} prefixes, so the penalty
per eviction is lower, and the headroom for improvement is correspondingly reduced. 

\subsection{Predictor Accuracy}

\begin{figure}[t]
  \centering
  \begin{subfigure}{\columnwidth}
    \centering
    \includegraphics[width=\linewidth]{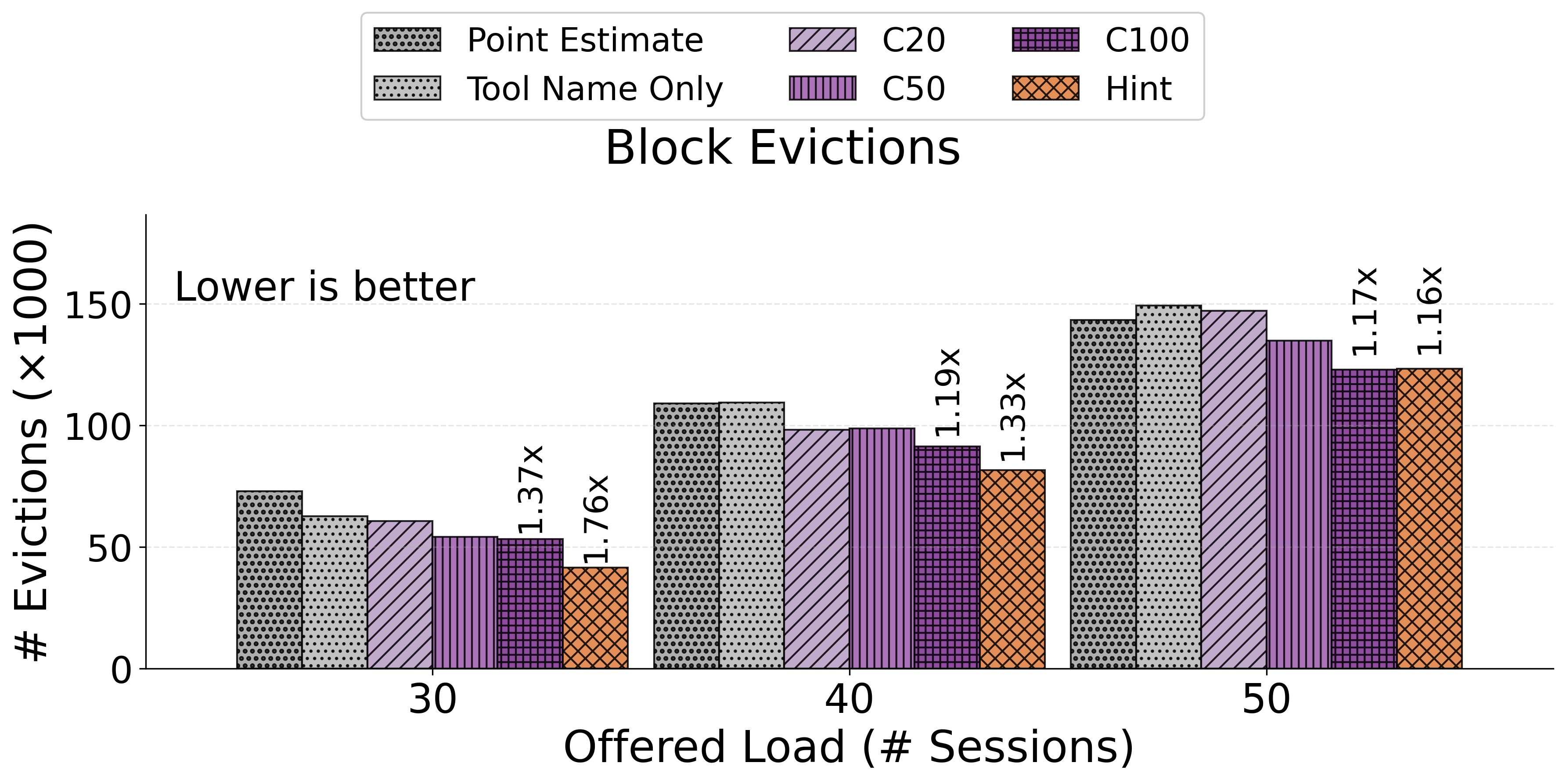}
    \caption{Block Evictions.}
    \label{fig:predictor-accuracy-evictions}
  \end{subfigure}
  \begin{subfigure}{\columnwidth}
    \centering
    \includegraphics[width=\linewidth]{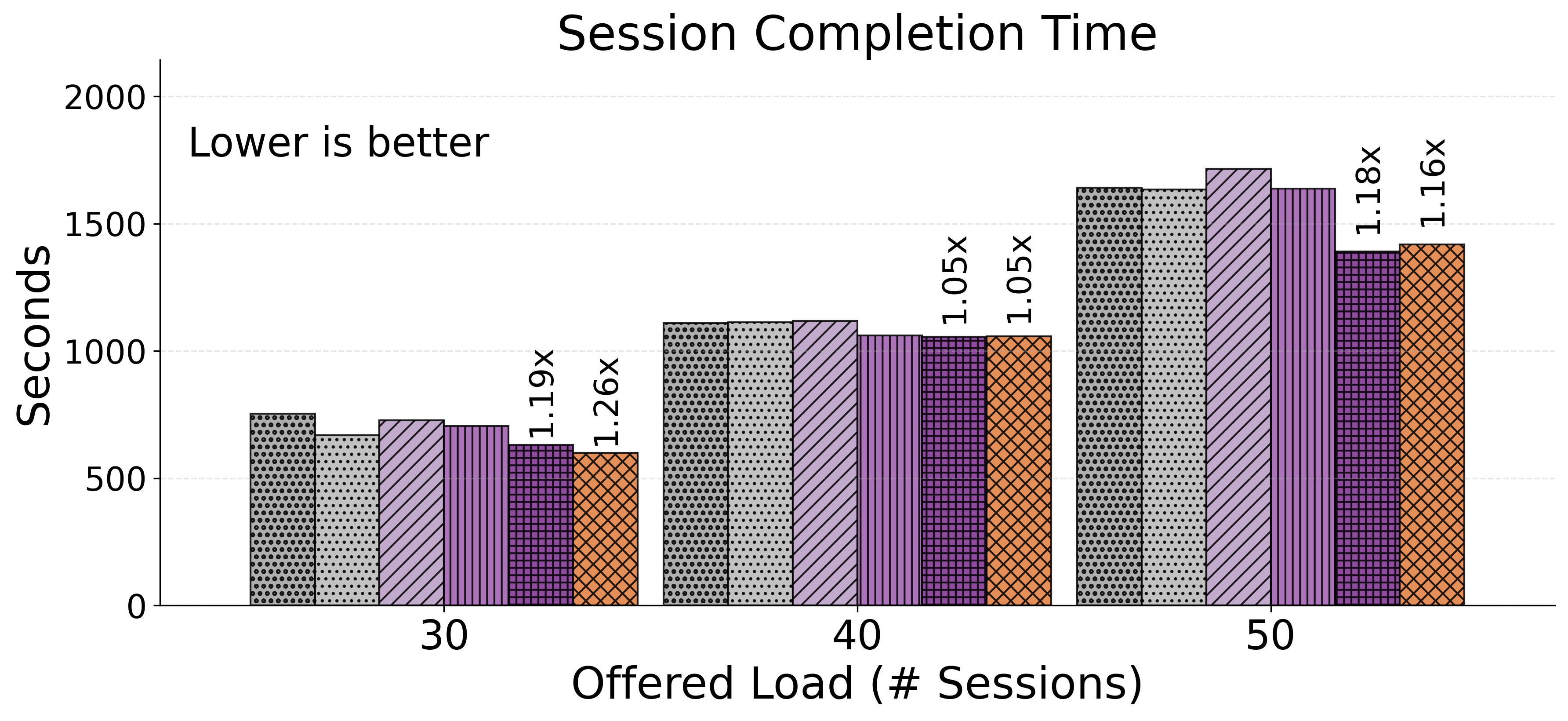}
    \caption{Mean session completion time.}
    \label{fig:predictor-accuracy-session}
  \end{subfigure}
\caption{Eviction count and session completion time in seconds. Lower values are better for both. \textbf{CX}=Tool argument based clustering with X clusters. \textbf{Hint}=Trace replay with ground truth tool durations.}
  \label{fig:eval-predictor-accuracy}
\end{figure}

We evaluate how prediction granularity affects eviction count and
session completion time by progressively refining the estimator and
measuring the resulting system efficiency.
We consider the following predictors: 
\textbf{Point Estimates} uses a single point estimate of $\hat{\tau}$ derived from the global mean tool duration across all tool types and sessions. 
\textbf{Tool Name Only} conditions the estimate on tool name alone, using the per-tool empirical duration distribution (\S\ref{sec:design}).
\textbf{C20}, \textbf{C50}, and \textbf{C100} further refine the
estimate by clustering tool calls based on tool arguments,
using 20, 50, and 100 clusters respectively. Increasing the cluster count beyond 100 yielded no additional clusters, as the data did not 
support finer granularity. Using cluster \textbf{Hint} uses
ground-truth tool duration from the trace. 
Figure~\ref{fig:eval-predictor-accuracy} reports eviction counts and session completion times across four predictor variants and three load levels ($N=\{30,40,50\}$). We observe that as the prediction granularity increases from \textbf{Point Estimate} to \textbf{C100},
both eviction count and session completion time decrease across values of N. \textbf{C100} achieves the highest performance, with upto $19\%$ improvement in session completion time. 

\subsection{Scheduling Overheads}
\begin{figure}
    \centering
    \includegraphics[width=1\linewidth]{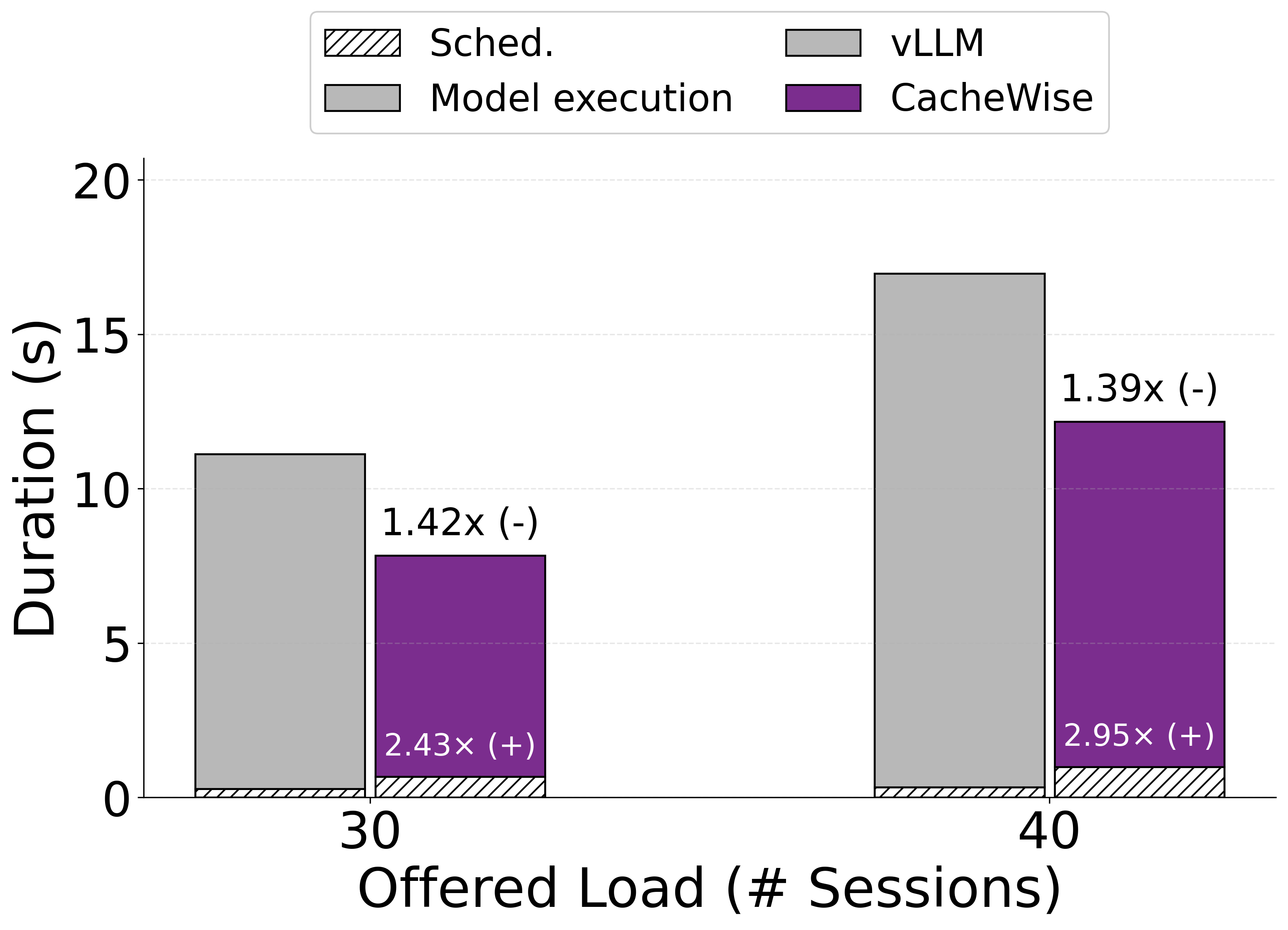}
    \caption{CPU scheduling overheads (Sched.) vs. \gpu{} execution overhead (Model Exec.) for different systems.}
    \label{fig:sched-overhead}
\end{figure}

Figure~\ref{fig:sched-overhead} compares the CPU-side scheduling overhead incurred by the {\projectname}'s {\kvcache} management policies against the \gpu{} model execution time. The scheduling overhead corresponds to the time spent making scheduling and eviction decisions prior to each model iteration that executes on the \gpu{}. {\projectname} incurs \textasciitilde2.4$\times$--3$\times$ higher CPU scheduling overhead as those incurred by vLLM. However, the absolute CPU scheduling overheads are negligible when compared to the corresponding gains in lower model execution times. 
At $N=40$, {\projectname} incurs higher CPU scheduling overhead, increasing from 0.33\,s (vLLM) to 0.99\,s  (3.0$\times$ increase). However, this overhead increase is more than offset by a significant reduction in \gpu{} model execution time, which decreases from 16.6\,s to 11.2\,s (1.48$\times$ improvement). Overall, {\projectname} yields a net reduction of \textasciitilde4.7\,s per request, demonstrating that savings from reduced model execution far outweigh the additional CPU scheduling cost. In relative terms, scheduling overhead accounts for roughly \textasciitilde6\% of total request time under vLLM and increases to about \textasciitilde9\% under {\projectname}. However the absolute reduction in model execution time (\textasciitilde4.7\,s for $N=40$) is responsible for the significant gains in end-to-end performance.

\section{Related Work}

\vspace{0.05in}
\noindent\textbf{Serving systems for emerging LLM applications:} Our work differs from recent works, e.g., Pie~\cite{gim2025pie}, Autellix~\cite{autellix}, which create novel programming abstractions for agentic workloads. 
These systems tightly couple the agent implementation with the serving system and cannot work with existing widely deployed coding agents, while {\projectname} can because it only modifies the serving system for these workload-specific characteristics.

\vspace{0.05in}
\noindent\textbf{Session-aware vs.\ request-aware scheduling:}
Existing serving systems including vLLM~\cite{vllm} and
InferCept~\cite{infercept}, typically schedule each request independently rather than optimizing for session completion times. Although SGLang~\cite{sglang} introduces prefix-aware scheduling to reuse {\kvcache} across requests that share a common system prompt, their work primarily targets sharing across independent requests, while our work demonstrates this is especially valuable for long-running, closed-loop coding agent sessions.

\vspace{0.05in}
\noindent\textbf{Predictive vs.\ recency-based eviction:}
vLLM, SGLang, and Mooncake~\cite{mooncake} all rely on LRU eviction, which uses only past access
recency and is oblivious to future reuse.
InferCept~\cite{infercept} takes a step toward
prediction by estimating tool call durations from a
session-local moving average; however, this estimate is agnostic to tool type and arguments, limiting its accuracy for \codingagent{} workloads where tool durations vary by orders of magnitude across tool types \xref{sec:design}). InfiniGen~\cite{lee2024infinigen} dynamically manages {\kvcache} with speculative offloading but does not predict reuse from session metadata. Although predictive caching has been studied theoretically previously~\cite{lykouris2021competitive}, \projectname{} applies this practically for {\kvcache} management.

\vspace{0.05in}
\noindent\textbf{\kvcache{} tiered storage and memory management:}
LMCache~\cite{lmcache} and Mooncake~\cite{mooncake}
address memory pressure by offloading evicted {\kvcache} blocks to lower storage tiers (DRAM, SSD, or remote memory). NEO~\cite{jiang2025neo} offloads part of the attention computation and {\kvcache} state to the CPU to increase effective batch size.
Jenga~\cite{zhang2025jenga} introduces a two-level memory allocator for heterogeneous {\kvcache} embeddings, addressing memory fragmentation through LCM-based allocation.
EvicPress~\cite{feng2025evicpress} jointly applies lossy compression and adaptive eviction across storage tiers. Tiered storage is complementary to \projectname{}'s approach. By reducing the frequency of evictions, \projectname{} directly reduces the volume of data that must be moved across tiers, lowering the bandwidth and latency overhead that tiered systems must absorb. 


\section{Conclusion}

Coding agents are emerging as an important class of LLM workloads, but they differ fundamentally from traditional chat workloads. Through a study of CATraces, our collected dataset of real-world coding-assistant sessions, we show that coding agents execute as long-running closed loops, generate predominantly tool-initiated turns, and repeatedly reuse large, growing prefixes. These properties make serving efficiency depend not only on per-request execution, but on how well the system preserves and reuses {\kvcache} across turns within a session.

We present {\projectname}, a KVCache management layer for coding-agent workloads. {\projectname} combines prefix-aware scheduling, which prioritizes requests that can reuse resident {\kvcache}, with predictive eviction, which retains prefixes expected to be reused soon based on tool execution metadata. Implemented in vLLM and evaluated on real coding-agent traces, {\projectname} reduces {\kvcache} evictions by up to $2$--$2.6\times$ and improves end-to-end session completion time by up to \textasciitilde$3.5\times$.


\bibliographystyle{ACM-Reference-Format}
\bibliography{references}

@misc{toucan,
      title={TOUCAN: Synthesizing 1.5M Tool-Agentic Data from Real-World MCP Environments}, 
      author={Zhangchen Xu and Adriana Meza Soria and Shawn Tan and Anurag Roy and Ashish Sunil Agrawal and Radha Poovendran and Rameswar Panda},
      year={2025},
      eprint={2510.01179},
      archivePrefix={arXiv},
      primaryClass={cs.LG},
      url={https://arxiv.org/abs/2510.01179}, 
}

@misc{T1,
      title={T1: A Tool-Oriented Conversational Dataset for Multi-Turn Agentic Planning}, 
      author={Amartya Chakraborty and Paresh Dashore and Nadia Bathaee and Anmol Jain and Anirban Das and Shi-Xiong Zhang and Sambit Sahu and Milind Naphade and Genta Indra Winata},
      year={2025},
      eprint={2505.16986},
      archivePrefix={arXiv},
      primaryClass={cs.CL},
      url={https://arxiv.org/abs/2505.16986}, 
}

@misc{sharegpt52k,
  title        = {ShareGPT 52K Dataset},
  author       = {RyokoAI},
  year         = {2023},
  howpublished = {\url{https://huggingface.co/datasets/RyokoAI/ShareGPT52K}},
  note         = {Accessed: 2026-03-11}
}

@misc{swe-smith,
      title={SWE-smith: Scaling Data for Software Engineering Agents}, 
      author={John Yang and Kilian Lieret and Carlos E. Jimenez and Alexander Wettig and Kabir Khandpur and Yanzhe Zhang and Binyuan Hui and Ofir Press and Ludwig Schmidt and Diyi Yang},
      year={2025},
      eprint={2504.21798},
      archivePrefix={arXiv},
      primaryClass={cs.SE},
      url={https://arxiv.org/abs/2504.21798}, 
}

@online{bentoml_prefix_aware_routing,
  title  = {Prefix-aware Routing},
  author = {{BentoML}},
  year   = {2025},
  url    = {https://bentoml.com/llm/inference-optimization/prefix-aware-routing},
  urldate = {2026-03-12}
}

@misc{linkedin_prefix_aware_routing,
      title={LLM Query Scheduling with Prefix Reuse and Latency Constraints}, 
      author={Gregory Dexter and Shao Tang and Ata Fatahi Baarzi and Qingquan Song and Tejas Dharamsi and Aman Gupta},
      year={2025},
      eprint={2502.04677},
      archivePrefix={arXiv},
      primaryClass={cs.DS},
      url={https://arxiv.org/abs/2502.04677}, 
}

@misc{vllm_prefixaware_routing,
  title        = {Prefix Aware Routing — vLLM Production Stack},
  howpublished = {\url{https://docs.vllm.ai/projects/production-stack/en/vllm-stack-0.1.4/tutorials/prefixaware.html}},
  note         = {Accessed: 2026-03-11},
  year         = {2025}
}

@inproceedings{
  infercept,
  title={INFERCEPT: Efficient Intercept Support for Augmented Large Language Model
Inference},
  author={Reyna Abhyankar and Zijian He and Vikranth Srivatsa and Hao Zhang and Yiying Zhang},
  booktitle={Forty-first International Conference on Machine Learning},
  year={2024},
  month=Jul,
  address={Vienna, Austria},
}

@misc{transformers,
      title={Attention Is All You Need}, 
      author={Ashish Vaswani and Noam Shazeer and Niki Parmar and Jakob Uszkoreit and Llion Jones and Aidan N. Gomez and Lukasz Kaiser and Illia Polosukhin},
      year={2023},
      eprint={1706.03762},
      archivePrefix={arXiv},
      primaryClass={cs.CL},
      url={https://arxiv.org/abs/1706.03762}, 
}

@inproceedings {orca,
author = {Gyeong-In Yu and Joo Seong Jeong and Geon-Woo Kim and Soojeong Kim and Byung-Gon Chun},
title = {Orca: A Distributed Serving System for {Transformer-Based} Generative Models},
booktitle = {16th USENIX Symposium on Operating Systems Design and Implementation (OSDI 22)},
year = {2022},
isbn = {978-1-939133-28-1},
address = {Carlsbad, CA},
pages = {521--538},
url = {https://www.usenix.org/conference/osdi22/presentation/yu},
publisher = {USENIX Association},
month = jul
}

@inproceedings{vllm,
author = {Kwon, Woosuk and Li, Zhuohan and Zhuang, Siyuan and Sheng, Ying and Zheng, Lianmin and Yu, Cody Hao and Gonzalez, Joseph and Zhang, Hao and Stoica, Ion},
title = {Efficient Memory Management for Large Language Model Serving with PagedAttention},
year = {2023},
isbn = {9798400702297},
publisher = {Association for Computing Machinery},
address = {New York, NY, USA},
url = {https://doi.org/10.1145/3600006.3613165},
doi = {10.1145/3600006.3613165},
booktitle = {Proceedings of the 29th Symposium on Operating Systems Principles},
pages = {611–626},
numpages = {16},
location = {Koblenz, Germany},
series = {SOSP '23}
}

@misc{sglang,
      title={SGLang: Efficient Execution of Structured Language Model Programs}, 
      author={Lianmin Zheng and Liangsheng Yin and Zhiqiang Xie and Chuyue Sun and Jeff Huang and Cody Hao Yu and Shiyi Cao and Christos Kozyrakis and Ion Stoica and Joseph E. Gonzalez and Clark Barrett and Ying Sheng},
      year={2024},
      eprint={2312.07104},
      archivePrefix={arXiv},
      primaryClass={cs.AI},
      url={https://arxiv.org/abs/2312.07104}, 
}

@misc{lmcache,
      title={LMCache: An Efficient KV Cache Layer for Enterprise-Scale LLM Inference}, 
      author={Yihua Cheng and Yuhan Liu and Jiayi Yao and Yuwei An and Xiaokun Chen and Shaoting Feng and Yuyang Huang and Samuel Shen and Kuntai Du and Junchen Jiang},
      year={2025},
      eprint={2510.09665},
      archivePrefix={arXiv},
      primaryClass={cs.LG},
      url={https://arxiv.org/abs/2510.09665}, 
}

@inproceedings {mooncake,
author = {Ruoyu Qin and Zheming Li and Weiran He and Jialei Cui and Feng Ren and Mingxing Zhang and Yongwei Wu and Weimin Zheng and Xinran Xu},
title = {Mooncake: Trading More Storage for Less Computation {\textemdash} A {KVCache-centric} Architecture for Serving {LLM} Chatbot},
booktitle = {23rd USENIX Conference on File and Storage Technologies (FAST 25)},
year = {2025},
isbn = {978-1-939133-45-8},
address = {Santa Clara, CA},
pages = {155--170},
url = {https://www.usenix.org/conference/fast25/presentation/qin},
publisher = {USENIX Association},
month = feb
}

@misc{claude_code,
  author       = {Anthropic},
  title        = {Claude Code — AI coding assistant},
  howpublished = {\url{https://www.claude.com/product/claude-code}},
  note         = {Accessed: 2026-03-11},
  year         = {2025}
}

@inproceedings{distserve,
  author = {Yinmin Zhong and Shengyu Liu and Junda Chen and Jianbo Hu and Yibo Zhu and Xuanzhe Liu and Xin Jin and Hao Zhang},
  title = {{DistServe}: Disaggregating Prefill and Decoding for Goodput-optimized Large Language Model Serving},
  booktitle = {18th USENIX Symposium on Operating Systems Design and Implementation (OSDI 24)},
  pages = {193--210},
  year = {2024},
  publisher = {USENIX Association}
}

@inproceedings{sarathi-serve,
author = {Agrawal, Amey and Kedia, Nitin and Panwar, Ashish and Mohan, Jayashree and Kwatra, Nipun and Gulavani, Bhargav S. and Tumanov, Alexey and Ramjee, Ramachandran},
title = {Taming throughput-latency tradeoff in LLM inference with sarathi-serve},
year = {2024},
isbn = {978-1-939133-40-3},
publisher = {USENIX Association},
address = {USA},
booktitle = {Proceedings of the 18th USENIX Conference on Operating Systems Design and Implementation},
articleno = {7},
numpages = {18},
location = {Santa Clara, CA, USA},
series = {OSDI'24}
}

@misc{autellix,
      title={Autellix: An Efficient Serving Engine for LLM Agents as General Programs}, 
      author={Michael Luo and Xiaoxiang Shi and Colin Cai and Tianjun Zhang and Justin Wong and Yichuan Wang and Chi Wang and Yanping Huang and Zhifeng Chen and Joseph E. Gonzalez and Ion Stoica},
      year={2025},
      eprint={2502.13965},
      archivePrefix={arXiv},
      primaryClass={cs.LG},
      url={https://arxiv.org/abs/2502.13965}, 
}

@inproceedings{splitwise,
  author = {Pratyush Patel and Esha Choukse and Chaojie Zhang and \'{I}\~{n}igo Goiri and Aashaka Shah and Saeed Maleki and Ricardo Bianchini},
  title = {Splitwise: Efficient Generative {LLM} Inference Using Phase Splitting},
  booktitle = {Proceedings of the 51st Annual International Symposium on Computer Architecture (ISCA)},
  pages = {118--132},
  year = {2024},
  doi = {10.1109/ISCA59077.2024.00019}
}

@misc{qwen,
      title={Qwen Technical Report}, 
      author={Jinze Bai and Shuai Bai and Yunfei Chu and Zeyu Cui and Kai Dang and Xiaodong Deng and Yang Fan and Wenbin Ge and Yu Han and Fei Huang and Binyuan Hui and Luo Ji and Mei Li and Junyang Lin and Runji Lin and Dayiheng Liu and Gao Liu and Chengqiang Lu and Keming Lu and Jianxin Ma and Rui Men and Xingzhang Ren and Xuancheng Ren and Chuanqi Tan and Sinan Tan and Jianhong Tu and Peng Wang and Shijie Wang and Wei Wang and Shengguang Wu and Benfeng Xu and Jin Xu and An Yang and Hao Yang and Jian Yang and Shusheng Yang and Yang Yao and Bowen Yu and Hongyi Yuan and Zheng Yuan and Jianwei Zhang and Xingxuan Zhang and Yichang Zhang and Zhenru Zhang and Chang Zhou and Jingren Zhou and Xiaohuan Zhou and Tianhang Zhu},
      year={2023},
      eprint={2309.16609},
      archivePrefix={arXiv},
      primaryClass={cs.CL},
      url={https://arxiv.org/abs/2309.16609}, 
}

@misc{tfid_vector,
  title        = {{TfidfVectorizer}},
  author       = {{scikit-learn developers}},
  howpublished = {\url{https://scikit-learn.org/stable/modules/generated/sklearn.feature_extraction.text.TfidfVectorizer.html}},
  year         = {2024},
  note         = {Accessed: 2026-01-22}
}

@misc{openai_codex,
  title        = {{OpenAI Codex}},
  author       = {OpenAI},
  howpublished = {\url{https://github.com/openai/codex}},
  year         = {2023},
  note         = {GitHub repository, accessed January 24, 2026}
}

@misc{gorilla,
      title={Gorilla: Large Language Model Connected with Massive APIs}, 
      author={Shishir G. Patil and Tianjun Zhang and Xin Wang and Joseph E. Gonzalez},
      year={2023},
      eprint={2305.15334},
      archivePrefix={arXiv},
      primaryClass={cs.CL},
      url={https://arxiv.org/abs/2305.15334}, 
}

@misc{github_copilot,
  title        = {GitHub Copilot},
  author       = {{GitHub}},
  howpublished = {\url{https://github.com/features/copilot}},
  year         = {2021},
  note         = {AI-powered code completion tool}
}

@misc{default_chunked_prefill,
  title        = {vLLM Chunked Prefill},
  author       = {{vLLM}},
  howpublished = {\url{https://docs.vllm.ai/en/v0.4.2/models/performance.html}},
  year         = {2026},
  note         = {Default chunked-prefill size in vLLM}
}

@INPROCEEDINGS{lwl,
  author={Harchol-Balter, Mor and Scheller-Wolf, Alan and Young, Andrew},
  booktitle={2009 47th Annual Allerton Conference on Communication, Control, and Computing (Allerton)}, 
  title={Why segregating short jobs from long jobs under high variability is not always a win}, 
  year={2009},
  volume={},
  number={},
  pages={121-127},
  doi={10.1109/ALLERTON.2009.5394853}}

@ARTICLE{belady,
  author={Belady, L. A.},
  journal={IBM Systems Journal}, 
  title={A study of replacement algorithms for a virtual-storage computer}, 
  year={1966},
  volume={5},
  number={2},
  pages={78-101},
  doi={10.1147/sj.52.0078}}

@misc{tfidf_doc,
  title        = {Scikit-learn TF-IDF documentation},
  author       = {{scikit-learn}},
  howpublished = {\url{https://scikit-learn.org/stable/modules/generated/sklearn.feature_extraction.text.TfidfVectorizer.html}},
  year         = {2026},
  note         = {}
}

@inproceedings{gqa,
    title = "{GQA}: Training Generalized Multi-Query Transformer Models from Multi-Head Checkpoints",
    author = "Ainslie, Joshua  and
      Lee-Thorp, James  and
      de Jong, Michiel  and
      Zemlyanskiy, Yury  and
      Lebron, Federico  and
      Sanghai, Sumit",
    editor = "Bouamor, Houda  and
      Pino, Juan  and
      Bali, Kalika",
    booktitle = "Proceedings of the 2023 Conference on Empirical Methods in Natural Language Processing",
    month = dec,
    year = "2023",
    address = "Singapore",
    publisher = "Association for Computational Linguistics",
    url = "https://aclanthology.org/2023.emnlp-main.298/",
    doi = "10.18653/v1/2023.emnlp-main.298",
    pages = "4895--4901"
}

@misc{mqa,
      title={Fast Transformer Decoding: One Write-Head is All You Need}, 
      author={Noam Shazeer},
      year={2019},
      eprint={1911.02150},
      archivePrefix={arXiv},
      primaryClass={cs.NE},
      url={https://arxiv.org/abs/1911.02150}, 
}

@misc{megatron,
      title={Megatron-LM: Training Multi-Billion Parameter Language Models Using Model Parallelism}, 
      author={Mohammad Shoeybi and Mostofa Patwary and Raul Puri and Patrick LeGresley and Jared Casper and Bryan Catanzaro},
      year={2020},
      eprint={1909.08053},
      archivePrefix={arXiv},
      primaryClass={cs.CL},
      url={https://arxiv.org/abs/1909.08053}, 
}

@inproceedings{zhang2025jenga,
  author = {Chen Zhang and Kuntai Du and Shu Liu and Woosuk Kwon and Xiangxi Mo and Yufeng Wang and Xiaoxuan Liu and Kaichao You and Zhuohan Li and Mingsheng Long and Jidong Zhai and Joseph Gonzalez and Ion Stoica},
  title = {Jenga: Effective Memory Management for Serving {LLM} with Heterogeneity},
  booktitle = {Proceedings of the ACM SIGOPS 31st Symposium on Operating Systems Principles (SOSP)},
  pages = {446--461},
  year = {2025},
  doi = {10.1145/3731569.3764823}
}

@inproceedings{du2025prefillonly,
  author = {Kuntai Du and Bowen Wang and Chen Zhang and Yiming Cheng and Qing Lan and Hejian Sang and Yihua Cheng and Jiayi Yao and Xiaoxuan Liu and Yifan Qiao and Ion Stoica and Junchen Jiang},
  title = {{PrefillOnly}: An Inference Engine for Prefill-only Workloads in Large Language Model Applications},
  booktitle = {Proceedings of the ACM SIGOPS 31st Symposium on Operating Systems Principles (SOSP)},
  pages = {399--414},
  year = {2025},
  doi = {10.1145/3731569.3764834}
}

@inproceedings{gim2025pie,
  author = {In Gim and Zhiyao Ma and Seung-seob Lee and Lin Zhong},
  title = {Pie: A Programmable Serving System for Emerging {LLM} Applications},
  booktitle = {Proceedings of the ACM SIGOPS 31st Symposium on Operating Systems Principles (SOSP)},
  year = {2025},
  doi = {10.1145/3731569.3764814}
}

@inproceedings{lee2024infinigen,
  author = {Wonbeom Lee and Jungi Lee and Junghwan Seo and Jaewoong Sim},
  title = {{InfiniGen}: Efficient Generative Inference of Large Language Models with Dynamic {KV} Cache Management},
  booktitle = {18th USENIX Symposium on Operating Systems Design and Implementation (OSDI 24)},
  pages = {155--172},
  year = {2024},
  publisher = {USENIX Association}
}

@inproceedings{sun2024llumnix,
  author = {Biao Sun and Ziming Huang and Hanyu Zhao and Wencong Xiao and Xinyi Zhang and Yong Li and Wei Lin},
  title = {Llumnix: Dynamic Scheduling for Large Language Model Serving},
  booktitle = {18th USENIX Symposium on Operating Systems Design and Implementation (OSDI 24)},
  pages = {173--191},
  year = {2024},
  publisher = {USENIX Association}
}

@inproceedings{fu2024serverlessllm,
  author = {Yao Fu and Leyang Xue and Yeqi Huang and Andrei-Octavian Brabete and Dmitrii Ustiugov and Yuvraj Patel and Luo Mai},
  title = {{ServerlessLLM}: Low-Latency Serverless Inference for Large Language Models},
  booktitle = {18th USENIX Symposium on Operating Systems Design and Implementation (OSDI 24)},
  pages = {135--153},
  year = {2024},
  publisher = {USENIX Association}
}

@inproceedings{jiang2025neo,
  author = {Xuanlin Jiang and Yang Zhou and Shiyi Cao and Ion Stoica and Minlan Yu},
  title = {{NEO}: Saving {GPU} Memory Crisis with {CPU} Offloading for Online {LLM} Inference},
  booktitle = {Proceedings of Machine Learning and Systems (MLSys)},
  year = {2025}
}

@inproceedings{yao2023react,
  author = {Shunyu Yao and Jeffrey Zhao and Dian Yu and Nan Du and Izhak Shafran and Karthik R. Narasimhan and Yuan Cao},
  title = {{ReAct}: Synergizing Reasoning and Acting in Language Models},
  booktitle = {International Conference on Learning Representations (ICLR)},
  year = {2023}
}

@inproceedings{shinn2023reflexion,
  author = {Noah Shinn and Federico Cassano and Ashwin Gopinath and Karthik Narasimhan and Shunyu Yao},
  title = {Reflexion: Language Agents with Verbal Reinforcement Learning},
  booktitle = {Advances in Neural Information Processing Systems (NeurIPS)},
  year = {2023}
}

@inproceedings{jimenez2024swebench,
  author = {Carlos E. Jimenez and John Yang and Alexander Wettig and Shunyu Yao and Kexin Pei and Ofir Press and Karthik R. Narasimhan},
  title = {{SWE-bench}: Can Language Models Resolve Real-world {GitHub} Issues?},
  booktitle = {International Conference on Learning Representations (ICLR)},
  year = {2024}
}

@inproceedings{yang2024sweagent,
  author = {John Yang and Carlos E. Jimenez and Alexander Wettig and Kilian Lieret and Shunyu Yao and Karthik R. Narasimhan and Ofir Press},
  title = {{SWE-agent}: Agent-Computer Interfaces Enable Automated Software Engineering},
  booktitle = {Advances in Neural Information Processing Systems (NeurIPS)},
  year = {2024}
}

@inproceedings{zhang2024autocoderover,
  author = {Yuntong Zhang and Haifeng Ruan and Zhiyu Fan and Abhik Roychoudhury},
  title = {{AutoCodeRover}: Autonomous Program Improvement},
  booktitle = {Proceedings of the 33rd ACM SIGSOFT International Symposium on Software Testing and Analysis (ISSTA)},
  pages = {1592--1604},
  year = {2024},
  doi = {10.1145/3650212.3680384}
}

@article{lykouris2021competitive,
  author = {Thodoris Lykouris and Sergei Vassilvitskii},
  title = {Competitive Caching with Machine Learned Advice},
  journal = {Journal of the ACM},
  volume = {68},
  number = {4},
  pages = {1--25},
  year = {2021},
  doi = {10.1145/3447579}
}

@article{feng2025evicpress,
  author = {Shaoting Feng and Yuhan Liu and Hanchen Li and Xiaokun Chen and Samuel Shen and Kuntai Du and Zhuohan Gu and Rui Zhang and Yuyang Huang and Yihua Cheng and Jiayi Yao and Qizheng Zhang and Ganesh Ananthanarayanan and Junchen Jiang},
  title = {{EvicPress}: Joint {KV}-Cache Compression and Eviction for Efficient {LLM} Serving},
  journal = {arXiv preprint arXiv:2512.14946},
  year = {2025}
}

@misc{kmeans,
  title        = {KMeans Scikit Learn},
  author       = {{KMeans}},
  howpublished = {\url{https://scikit-learn.org/stable/modules/generated/sklearn.cluster.KMeans.html}},
  year         = {2026},
}

\end{document}